\documentclass[12pt]{article}
\usepackage[margin=1in]{geometry}
\usepackage{amsmath}
\usepackage{graphicx}
\usepackage{siunitx}
\usepackage{float}
\usepackage{array}
\usepackage[version=4]{mhchem}
\usepackage{hyperref}
\usepackage{url}
\usepackage{ragged2e}
\usepackage{setspace}
\usepackage{authblk}
\usepackage{tabularx}
\DeclareSIUnit\angstrom{\mathrm{Å}}
\DeclareSIUnit{\RIU}{RIU}
\usepackage{soul}
\usepackage{booktabs}
\usepackage[table]{xcolor}
\newcommand{\revcolor}{black}
\newcommand{\rev}[1]{\textcolor{\revcolor}{#1}}

\title{\textbf{Nonlinear Optical Properties of MXenes for Photonic Applications}}
\author{Alina Karabchevsky, Shadad Watad, Aviad Katiyi, Muhammad Sharif Uddin, Nicolás Abadía, Anupma Thakur, Babak Anasori}

\date{}

\begin{document}

\maketitle

\doublespacing

\section*{Abstract}
MXenes - a rapidly expanding family of two-dimensional transition-metal carbides, nitrides, and carbonitrides - have emerged as powerful nonlinear photonic materials owing to their metallic conductivity, ultrafast carrier dynamics, broadband optical response, and highly tunable surface chemistry. This Review provides the first comprehensive overview of the nonlinear optical (NLO) properties of MXenes and their growing impact on advanced photonic technologies. We discuss state-of-the-art synthesis strategies, from fluoride-based etching to molten-salt, electrochemical, and CVD-enabled routes, highlighting how structural quality and surface terminations govern optical behaviour. A detailed analysis of third-order nonlinearities including Kerr effects, two- and multi-photon absorption, saturable and reverse-saturable absorption, and harmonic generation is presented, supported by quantitative comparisons across MXene families. We further examine modelling approaches linking band structure to nonlinear response and survey application frontiers such as gas sensing, ultrafast lasers, all-optical limiters, and emerging photonic neural networks. Finally, we outline future research directions spanning material engineering, device integration, and quantum/nonlinear metasurface platforms. Together, these insights position MXenes as a transformative materials platform for next-generation nonlinear photonics and ultrafast all-optical information processing.\\
\textbf{Keywords:} MXenes, Optical Kerr effect, 2D materials, Nonlinear optics

\section{Introduction to MXenes}
Nonlinear optical (NLO) materials are fundamental to modern photonic technologies, enabling ultrafast signal processing, frequency conversion, and all-optical information manipulation in applications ranging from tunable lasers and programmable photonic circuits to quantum optics and optical communications ~\cite{ref:duarte2017tunable, ref:bogaerts2020programmable, ref:agarwal2012quantum, ref:agrawal2012fiber}. Despite their considerable potential, the intrinsically weak nonlinear response of conventional materials remains a major bottleneck in realizing compact, energy-efficient, and high-speed photonic devices. This challenge has driven the exploration of alternative material platforms with enhanced nonlinear susceptibilities and stronger light–matter interactions.

Two-dimensional (2D) materials have emerged as promising candidates for nonlinear photonics due to their reduced dimensionality, strong optical confinement, and tunable electronic structures~\cite{ref:kumbhakar2021advance, ref:adeshina2025exploring}. Among them, graphene has been investigated for nonlinear optical applications, including frequency conversion and plasmonic enhancement~\cite{ref:hamed2020graphene, ref:karimi2023plasmon}. Despite its exceptional carrier mobility and broadband response, graphene suffers from intrinsic limitations, including its zero bandgap and limited modulation depth, which constrain its effectiveness in tunable nonlinear photonic devices. To overcome these challenges, a wide range of alternative 2D materials, including transition metal dichalcogenides (TMDs), black phosphorus, and topological insulators, have been explored. While these materials exhibit strong nonlinear responses, their practical deployment is often constrained by synthesis complexity, limited stability, or difficulty achieving large-area, high-quality monolayer films, particularly via techniques such as chemical vapor deposition (CVD).

\begin{figure}[htbp]
    \centering
    \includegraphics[width=0.95\textwidth]{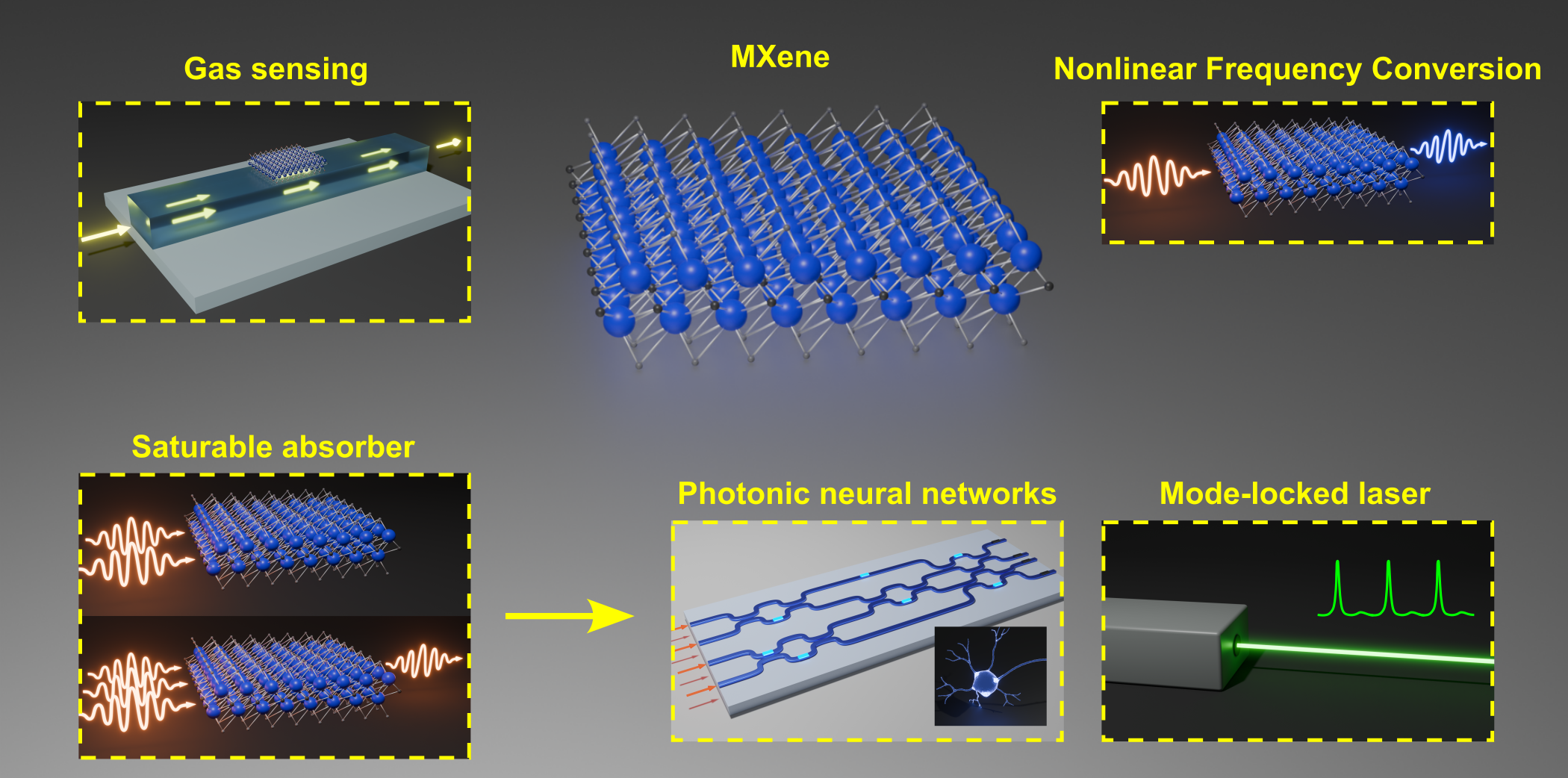}
    \caption{Artistic impression of the nonlinear optical landscape of MXenes.}
    \label{fig:NLO_intro}
\end{figure}

Since their discovery in 2011~\cite{ref:barsoum2011elastic}, MXenes have attracted significant attention as promising materials for photonic applications. Their metallic-like electronic structure, characterized by a high density of free carriers, enables efficient interaction with optical fields across a broad spectral range. In addition, the presence of chemically tunable surface terminations provides a unique degree of control over their electronic and optical properties. These features collectively make MXenes particularly attractive for enhancing nonlinear optical interactions in emerging photonic platforms~\cite{ref:simon2025tailoring}. Unlike many other 2D materials, MXenes mainly have a centrosymmetric structure, and their nonlinear optical response is therefore governed by third-order susceptibility ($\chi^{(3)}$)~\cite{ref:minassian2025nonlinear}. This leads to pronounced nonlinear phenomena such as the optical Kerr effect, two-photon absorption (TPA), saturable (SA) and reverse-saturable absorption (RSA), and harmonic generation. These processes play a central role in enabling advanced photonic functionalities where nonlinear effects such as second-harmonic generation (SHG) and SA are integrated within waveguide-based photonic architectures. Importantly, the magnitude and sign of these nonlinearities can be engineered through composition, thickness, and surface termination, providing an unprecedented level of control over their optical response. MXenes can therefore be regarded as a $\chi^{(3)}$-dominant nonlinear photonic platform for next-generation photonic technologies. The nonlinear optical landscape of MXenes and its connection to emerging photonic functionalities is conceptualised in Fig. \ref{fig:NLO_intro}, linking their nonlinear response to representative emerging functionalities including nonlinear frequency conversion, saturable absorption, ultrafast lasers, optical sensing and photonic neural networks. This review examines the nonlinear optical properties of MXenes, explaining the relationship between their structure, electronic properties, and nonlinear response, and highlighting their potential for integration into emerging photonic applications.

\section{Synthesis Methods}
\begin{figure}[htbp]
    \centering
    \includegraphics[width=0.95\textwidth]{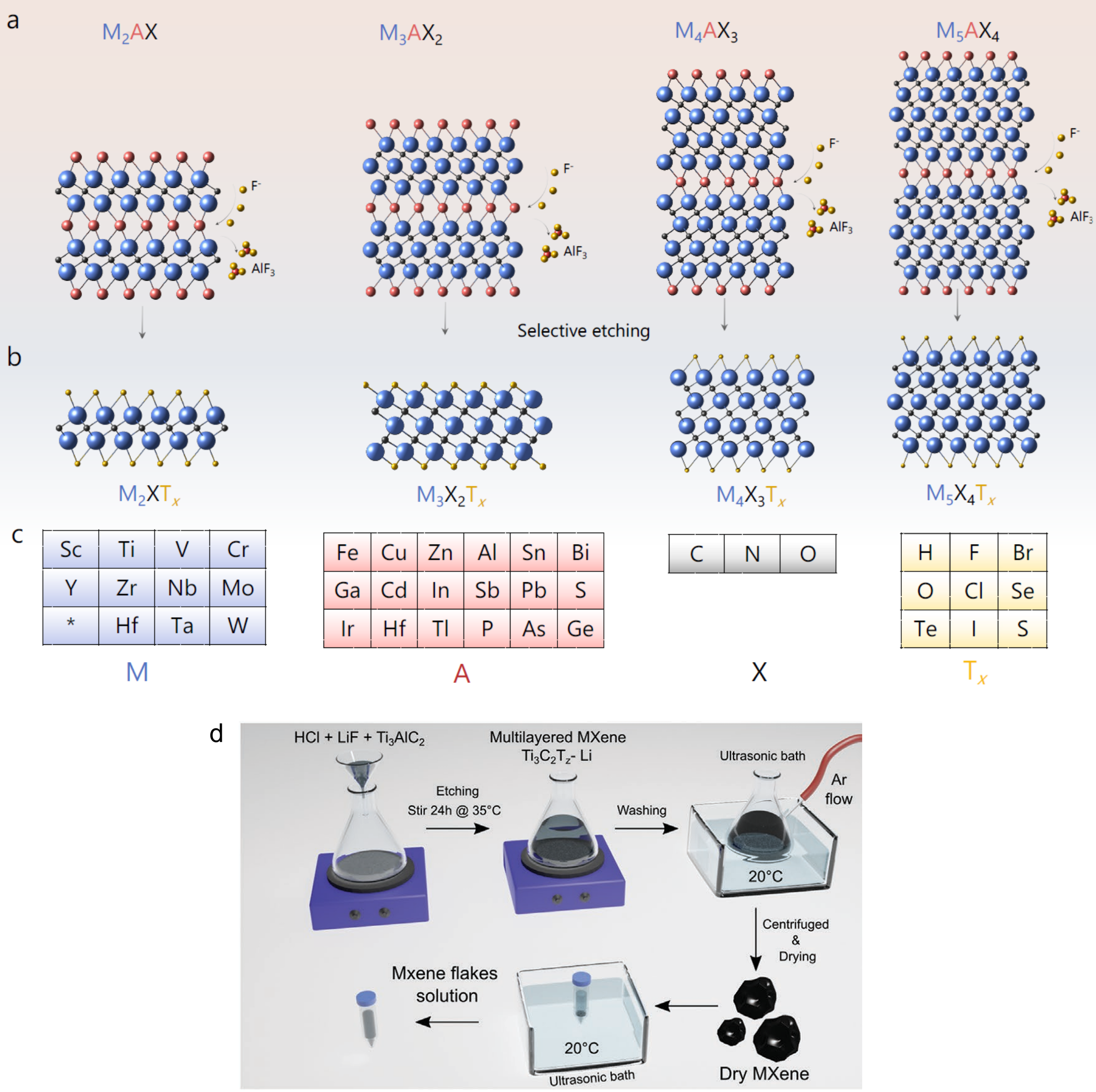}
    \caption{(a) Schematic of four types of MAX phases such as \ce{M2AX}, \ce{M3AX2}, \ce{M4AX3}, and \ce{M5AX4}. (b) MXenes are synthesized after selective etching of A layers (red atoms) and the formation of surface terminations (green atoms). (c) Elements used to synthesize MAX and MXenes to date (blue background are only reported as M layers, red background are the A elements, grey background represent the X elements, and \ce{Tx} represent surface terminations). The * in elements reported as M layers includes lanthanides such as Sm, Ce, Pr, Nd, Gd, Tb, Dy, Ho, Er, Yb, and Tm. (d) Illustration of the synthesis process of MXene (reproduced from~\cite{ref:hazan2023mxene}).}
    \label{fig:Fig2}
\end{figure}

MXenes, a family of 2D transition-metal carbides, nitrides, and carbonitrides, are mostly synthesized via a top-down synthesis approach by selective etching of A layers from their MAX phase precursors~\cite{ref:radovic2013max, ref:barsoum2000mn, ref:barsoum2011elastic}. MAX phases are layered carbides and nitrides that are known as machinable ceramics and represented by the chemical formula \ce{M_{n+1}AX_n}, where M is an early transition metal (blue in Fig.~\ref{fig:Fig2}a), A represents an A group element (red in Fig. \ref{fig:Fig2}a), and X represents carbon, or nitrogen (grey in Fig.~\ref{fig:Fig2}a) ~\cite{ref:sokol2019chemical, ref:tallman2013critical}.

In the MAX phase layered structure, the \ce{M_{n+1}X_n} layers are interleaved with the A layers. The key to successful MXene synthesis lies in the selective etching of the A-layer, breaking the relatively weaker M-A bonds compared to the stronger M-X bonds~\cite{ref:lim2022fundamentals}. Wet-chemical etching routes in fluoride-containing acidic solutions are the most commonly used methods for synthesizing MXenes. As MXenes are synthesized via the top-down selective etching process, the \ce{M_{n+1}X_n} compositions, their phase purity, flake morphology, and MXenes’ properties are dependent on the quality of the precursor MAX phases. Fig.~\ref{fig:Fig2}b shows the synthesized MXenes from their corresponding MAX phases, the \ce{M2AX}, \ce{M3AX2}, and \ce{M4AX3}. The A-layers are selectively etched in aqueous media using fluorine-containing acids, such as hydrofluoric acid, as shown in Fig.~\ref{fig:Fig2}a in yellow. After the selective removal of the A-layers, the outer transition metal (M) layers in \ce{M_{n+1}X_n} are commonly terminated by \ce{-F}, \ce{-O}, \ce{-Cl}, or \ce{-(OH)}, which are shown as \ce{T_x} in the MXenes’ \ce{M_{n+1}X_nT_x} formula~\cite{ref:lim2022fundamentals}. The presence of these functional groups on the basal planes of MXenes (the top and bottom exposed M layers) makes them highly hydrophilic with high negative surface charges (-30 to -60 mV)~\cite{ref:maleski2017dispersions}.

The strength of the M–A bonds in the precursor MAX phase determines the etching conditions required for MXene synthesis~\cite{ref:khaledialidusti2021high}. Some transition metals, such as chromium and molybdenum, also tend to create stronger M-A bonds relative to titanium~\cite{ref:lim2022fundamentals}. These variations in bond strength necessitate a range of synthesis methods to synthesize MXenes from different MAX phase precursors.

For MAX phases that are resistant to conventional aqueous acid etching, more recent methods such as molten salt etching have been investigated~\cite{ref:zhang2024delamination, ref:arole2021water, ref:li2020general, ref:li2019element}. For example, selective etching of \ce{Ti3SiC2} using \ce{CuCl2} molten salt to produce \ce{Ti3C2T_x} MXene~\cite{ref:li2020general}. The reaction breaks the strong M-A bond in \ce{Ti3SiC2}, selectively etching away the Si layers and leaving a multilayered \ce{Ti3C2T_x} MXene. This synthesis approach offers greater control over surface terminations and adds new Tx homogeneously on the MXene surface. For example, MXenes with Tx: \ce{-Se}, \ce{-Cl}, \ce{-Br}, or \ce{-I} are reported by molten salt etching methods~\cite{ref:arole2021water, ref:kamysbayev2020covalent}. This versatility in surface modification is critical in tailoring the MXene properties for diverse applications, such as chemical sensing~\cite{ref:liu2023bacterial, ref:ding2023chemical}.

Chemical vapor deposition (CVD) has also been reported for synthesizing MXenes~\cite{ref:wang2023direct, ref:xu2015large}. This approach allows for the direct synthesis of nitride MXenes, which have been challenging to produce using traditional etching methods. CVD offers improved control over the final composition and morphology of the resultant MXenes. The direct synthesis routes have also shown potential in the synthesis of MXenes. The first direct synthesis of \ce{Ti2CCl2} MXene in a high-temperature reaction of Ti, graphite, and \ce{TiCl4} at 950 °C in a sealed quartz ampoule has recently been reported~\cite{ref:wang2023direct}. Both in CVD and direct synthesis, MXenes can be synthesized directly, without the synthesis of MAX phases, making the process potentially simpler and enabling new MXene compositions.

Electrochemical etching is another synthesis method reported for MXenes. It is an emerging safer alternative to traditional HF-based etching processes~\cite{ref:chan2024fabrication}. The electrochemical etching has been used in the synthesis of \ce{Ti3C2} and \ce{Ti3CN} MXenes using an aqueous solution of tetrafluoroboric acid (\ce{HBF4}) as the electrolyte~\cite{ref:chan2024fabrication}. This approach involves a very low concentration of HF, making it less hazardous than conventional methods. Other synthesis methods, such as the hydrothermal method, have also been explored for MXene synthesis~\cite{ref:amrillah2022towards}. The choice of the synthesis method tailors the properties of the resulting MXenes. The harsh etching conditions, typically needed for MAX phases with stronger M-A bonds, can lead to introducing atomic-level defects in the MXenes~\cite{ref:wyatt2024alkali}. These atomic-scale vacancies and defects affect the electronic, thermal, mechanical, catalytic, and chemical properties of MXenes.

Fig.~\ref{fig:Fig2}c represents the elements that have been reported as M, A, X, and Tx structural design. The lanthanides are the latest addition to the MAX phase family that opens the possibility of synthesizing magnetic MAX phases~\cite{ref:tao2019atomically}. Furthermore, computational studies are also being employed to explore new synthesis pathways and guide experimental studies. Research efforts in this field are focused on designing novel, scalable, and sustainable synthesis methods for MXenes.

The structural diversity shown in Fig.~\ref{fig:Fig2} is chemically important, and also relevant to nonlinear photonics. Changing the MXene order from \ce{M2XT_x} to \ce{M5X4T_x} modifies the thickness of the transition-metal carbide/nitride slab, the density of electronic states, and the degree of metallic screening, thereby affecting the complex dielectric function, optical absorption, carrier dynamics, and third-order nonlinear susceptibility $\chi^{(3)}$~\cite{ref:shang2022nonlinear, ref:jiang2018broadband}. In parallel, the synthesis route controls surface terminations, defect density, flake size, and film morphology, which in turn determine the refractive index ($n$), extinction coefficient ($k$), carrier relaxation time ($\tau$), and nonlinear absorption coefficient ($\beta$) of the resulting MXene films~\cite{ref:li2020broadband, ref:wyatt2024alkali}. Beyond single-component MXenes, composite and heterostructure engineering provides an additional route to tailor nonlinear optical functionality. MXenes combined with polymers, transition-metal dichalcogenides, plasmonic nanostructures, or dielectric resonators can introduce interfacial charge transfer, enhanced light-matter interaction, local electromagnetic-field confinement, and modified ultrafast carrier dynamics. These effects can be used to tune saturable absorption, reverse saturable absorption, Kerr nonlinearities, and the effective $\chi^{(3)}$ response. Therefore, MXene synthesis should be viewed as a materials-design strategy for engineering nonlinear photonic platforms, including waveguide-integrated nonlinear activation elements, ultrafast modulators, optical limiters, and plasmonic or metasurface-enhanced nonlinear devices. A representative wet-chemical synthesis route for \ce{Ti3C2T_x} MXene, including etching, washing, and delamination, is illustrated in Figure ~\ref{fig:Fig2}d.

\section{Fundamental Nonlinear Properties}
NLO materials play a crucial role in all aspects of photonics, as they provide solutions for realizing ultrafast information processing through all-optical signal processing, photon generation, and optical memory. Recently, the nonlinear optical properties of 2D materials have attracted significant attention due to their superior performance compared to conventional bulk materials ~\cite{ref:novoselov20162d, ref:autere2018nonlinear}. The 2D materials, such as graphene~\cite{ref:sun2010graphene, ref:vermeulen2022perspectives, ref:gu2012regenerative}, TMDs~\cite{ref:wen2019nonlinear}, BP~\cite{ref:zheng2015characterization}, perovskite~\cite{ref:abdelwahab2019giant}, and MXene~\cite{ref:shang2022nonlinear, ref:hazan2023mxene, ref:jiang2018broadband} have been explored for their nonlinear properties like SA and SHG or THG and their broadband response. These materials show various intensity-dependent optical effects, including nonlinear absorption and refractive index modulation. These effects are key to ultrafast photonic functions. In the case of MXenes, such nonlinear behavior is mainly governed by strong third-order interactions, arising from their electronic structure and carrier dynamics, which enable efficient modulation of optical signals over a wide spectral range.

\subsection{Nonlinear Optical Properties}
MXenes exhibit nonlinear optical properties due to their unique electronic structure, surface chemistry, and 2D layered nature. Generally speaking, nonlinear optical responses in materials arise from the polarization $P(t)$ generated by an applied electric field $E(t)$, which can be expressed as a power series expansion as: 
\begin{equation}
    P(t) = \varepsilon_0 \left[ \chi^{(1)} E(t) + \chi^{(2)} E^2(t) + \chi^{(3)} E^3(t) + \cdots \right]
    \label{eq:polarization}
\end{equation}
where $\chi^{(n)}$ is the $n$-th order nonlinear susceptibility. The factors that fundamentally drive a strong $\chi^{(3)}$ response in MXenes are: free carrier density (Drude-like response), interband transitions, surface termination states, and plasmonic confinement~\cite{ref:fang2026band}. The first-order susceptibility $\chi^{(1)}$ is related to the linear optical properties, while higher orders describe the nonlinear effects. The second-order susceptibility $\chi^{(2)}$ is only non-zero in materials lacking inversion symmetry, which can't be found in the most centrosymmetric materials, including many 2D materials like MXenes. As a result, $\chi^{(2)}$ based phenomena, such as second-harmonic generation (SHG), sum-frequency generation (SFG), and difference-frequency generation (DFG), are limited or absent in pristine MXenes.

\begin{figure}[H]
    \centering
    \includegraphics[width=1\textwidth]{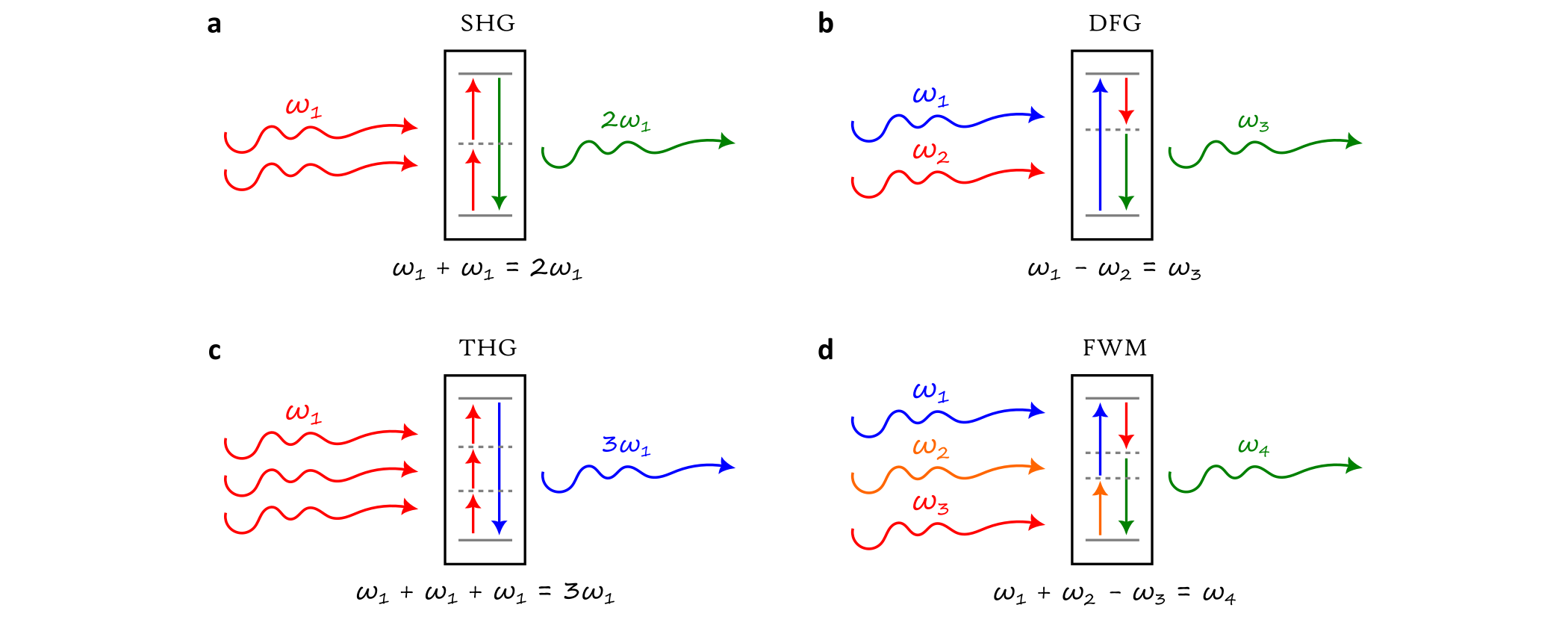}
    \caption{Schematic photon energy diagrams representing second-order $\chi^{(2)}$ and third-order $\chi^{(3)}$ nonlinear optical processes, including \rev{(a)}~second-harmonic generation (SHG), \rev{(b)}~difference-frequency generation (DFG), \rev{(c)}~third-harmonic generation (THG), and \rev{(d)}~four-wave mixing (FWM).} 
    \label{fig:nonlinear_processes}
\end{figure}

Third-order nonlinear susceptibility $\chi^{(3)}$, on the other hand, is present in all materials, regardless of symmetry, and dominates the nonlinear response in centrosymmetric materials. These processes include third-harmonic generation (THG), four-wave mixing (FWM), self-phase modulation (SPM), the optical Kerr effect, two-photon absorption (TPA), and saturable absorption (SA). \rev{The frequency-conversion processes are illustrated in Fig.~\ref{fig:nonlinear_processes}, namely the $\chi^{(2)}$ processes SHG and DFG (panels~a and~b), which are forbidden in centrosymmetric MXenes, and the $\chi^{(3)}$ processes THG and FWM (panels~c and~d), which are allowed regardless of symmetry.}

MXenes are typically centrosymmetric, and their nonlinear optical behavior is primarily attributed to strong third-order effects~\cite{ref:park2024ultrahigh, ref:zhang2021third}. Their high free carrier concentration, metallic-like conductivity, and surface-functional tunability make MXenes particularly promising for exploiting $\chi^{(3)}$ effects in photonic devices, including all-optical switches, modulators, and nonlinear neural network components. 

\subsubsection{Kerr effect}
MXenes exhibit strongly tunable Kerr nonlinearities, with reported nonlinear refractive index values ($n_2$) ranging from positive (self-focusing) to negative (self-defocusing), depending on their composition, excitation wavelength, and surface terminations (\ce{T_x}). This tunability makes MXenes promising candidates for photonic applications ranging from ultrafast modulators to optical limiters. The optical Kerr effect describes the underlying mechanism, a third-order nonlinear phenomenon in which the refractive index becomes dependent on the intensity of incident light. It can be expressed as: 
\begin{equation}
    n = n_0 + n_2 I
    \label{eq:kerr_effect}
\end{equation}
where $n_0$ is the linear refractive index, $n_2$ is the nonlinear refractive index, and $I$ is the optical intensity. The sign of $n_2$ determines the nature of the nonlinear response, with positive values corresponding to self-focusing and negative values to self-defocusing. These effects give rise to phenomena such as self-phase modulation (SPM), spectral broadening, and optical limiting in photonic devices.

\begin{figure}[htbp]
    \centering
    \includegraphics[width=1\textwidth]{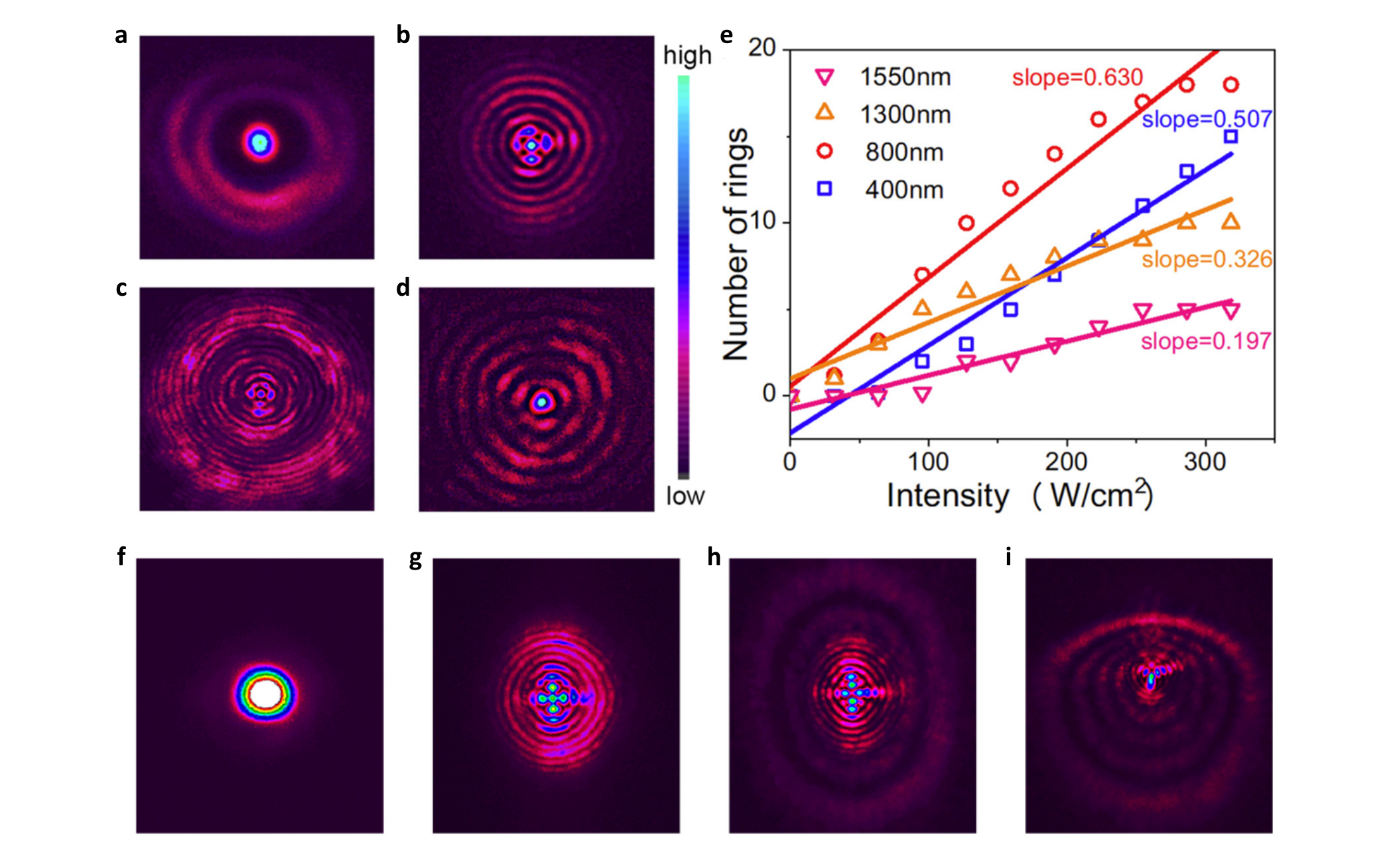}
    \caption{\rev{Spatial self-phase modulation in \ce{Nb2C} nanosheets. (a--d) Diffraction patterns at 1550, 1300, 800 and 400~nm, respectively; the color bar shows low to high intensity. (e) Number of diffraction rings versus incident intensity at the four wavelengths, with linear fits and their slopes. (f--i) Formation of the diffraction pattern at 1300~nm and $31.8~\mathrm{W/cm^2}$ after 0, 0.07, 0.53 and 1~s, respectively. Adapted from~\cite{ref:xiao2021near}.}}
    \label{fig:Fig4}
\end{figure}

Kerr nonlinearities in MXenes have been experimentally demonstrated across different families, revealing both self-defocusing and self-focusing responses. Among the earliest studies, Jiang et al.~\cite{ref:jiang2018broadband} reported negative $n_2$ values in \ce{Ti3C2T_x} MXene, ranging from $-4.66 \times10^{-16}$ to $-9.85 \times 10^{-16} ~\mathrm{cm^2/W}$ at wavelengths between \SI{800}{nm} and \SI{1800}{nm}, demonstrating pronounced self-defocusing behavior relevant to optical limiting and mid-infrared photonics. Li et al.~\cite{ref:li2020broadband} experimentally demonstrated broadband spatial self-phase modulation (SSPM) in \ce{Ti3C2T_x} over the spectral range from \SI{400}{nm} to approximately \SI{1000}{nm}, along with ultrafast carrier dynamics revealed by femtosecond transient absorption spectroscopy, highlighting its potential for ultrafast nonlinear photonic applications. Extending to other chemistries, \ce{V2C} exhibits a negative nonlinear refractive index of $-6.1 \times 10^{-10} ~\mathrm{cm^2/W}$ at \SI{1900}{nm} \cite{ref:lee2021investigation}. Out of \ce{V2C} family materials, \ce{V4C3} exhibits negative nonlinear refractive indices of $-4.8$ and $-3.5 \times 10^{-10} ~\mathrm{cm^2/W}$ at \SI{1560} and \SI{1910}{nm}, respectively~\cite{ref:lee2023nonlinear}. These results confirm their suitability for broadband modulation and mid-infrared photonic applications. In contrast, Nb-based MXenes generally exhibit positive Kerr nonlinearities. Xiao et al.~\cite{ref:xiao2021near} measured large $n_2$ values in \ce{Nb2C}, with values of $15.5,\, 13.1,$ and \, $9.53 \times 10^{-6} ~\mathrm{cm^2/W}$ at \SI{800}{nm}, \SI{1300}{nm}, and \SI{1550}{nm}, respectively, demonstrating strong self-focusing and efficient intensity-dependent phase modulation. \rev{These values were obtained from spatial self-phase modulation, where the number of diffraction rings observed at 1550, 1300, 800 and 400~nm (Fig.~\ref{fig:Fig4}a--d) increases linearly with the incident intensity (Fig.~\ref{fig:Fig4}e), and $n_2$ is extracted from the slope of each fit.} \rev{At a fixed wavelength and intensity, the pattern evolves in time, with the beam first self-focusing (Fig.~\ref{fig:Fig4}f), rings then appearing and multiplying (Fig.~\ref{fig:Fig4}g,h), and the pattern finally settling into a stable, asymmetric form (Fig.~\ref{fig:Fig4}i).} Interestingly, \ce{Nb4C3} was later found to exhibit a negative $n_2$ of $-1 \times 10^{-10} ~\mathrm{cm^2/W}$ at \SI{1900}{nm}~\cite{ref:lee2023investigation}, suggesting that stacking order and surface terminations also play critical roles. More recently, Stavrou et al.~\cite{ref:stavrou2026slow} investigated the Kerr nonlinearity of ordered \ce{Mo2Ti2C3T_x} MXene nanosheets in the telecommunication band using closed-aperture Z-scan measurements under femtosecond excitation, observing a self-defocusing nonlinear response and demonstrating that integration of \ce{Mo2Ti2C3T_x} within slow-light woodpile photonic crystals significantly enhanced the effective nonlinear refractive index, reaching values as high as $n_2\approx 5.5 \times 10^{-13} ~\mathrm{cm^2/W}$ and enabling ultrafast optical switching with a recovery time of approximately $2$ ps and switching rates approaching $500$ GHz.

These results demonstrate the strong tunability of Kerr nonlinearities in MXenes across different compositions and spectral ranges. Ti- and V-based MXenes predominantly exhibit negative $n_2$ values, making them suitable for optical limiting and broadband modulation ~\cite{ref:jiang2018broadband, ref:lee2021investigation, ref:lee2023nonlinear}, while Nb-based MXenes display large positive nonlinearities in the telecommunications range~\cite{ref:xiao2021near}, enabling efficient phase modulation. These findings highlight the potential of MXenes as a versatile platform for tailoring nonlinear optical responses through composition and surface engineering. Table \ref{tab:nonlinear_coeffs_n2} summarizes the reported nonlinear refractive index ($n_2$) values for representative MXene families across different wavelength ranges, highlighting the strong dependence of Kerr nonlinearity on material composition and excitation wavelength.

\begin{table}[htbp]
    \centering
    \caption{Summary of nonlinear refractive index $n_2$ values for different MXenes.}
    \rowcolors{2}{gray!8}{white}
    \begin{tabular}{cccc}
    \toprule
    \textbf{MXene} & \textbf{Wavelength (nm)} & \boldmath$n_2~(\mathrm{cm^2/W})$ & \textbf{Reference} \\
    \midrule
    \ce{Ti3C2T_x} & 800  & $-4.66 \times 10^{-16}$ & ~\cite{ref:jiang2018broadband} \\
    \ce{Ti3C2T_x} & 1064 & $-3.74 \times 10^{-16}$ & ~\cite{ref:jiang2018broadband} \\
    \ce{Ti3C2T_x} & 1550 & $-4.89 \times 10^{-16}$ & ~\cite{ref:jiang2018broadband} \\
    \ce{Ti3C2T_x} & 1800 & $-9.85 \times 10^{-16}$ & ~\cite{ref:jiang2018broadband} \\
    \ce{V2C}      & 1900 & $-6.1 \times 10^{-10}$  & ~\cite{ref:lee2021investigation} \\
    \ce{Nb2C}     & 800  & $15.5 \times 10^{-6}$   & ~\cite{ref:xiao2021near} \\
    \ce{Nb2C}     & 1300 & $13.1 \times 10^{-6}$   & ~\cite{ref:xiao2021near} \\
    \ce{Nb2C}     & 1550 & $9.53 \times 10^{-6}$   & ~\cite{ref:xiao2021near} \\
    \ce{V4C3}     & 1559 & $-4.8 \times 10^{-10}$  & ~\cite{ref:lee2023nonlinear} \\
    \ce{V4C3}     & 1911 & $-3.5 \times 10^{-10}$  & ~\cite{ref:lee2023nonlinear} \\
    \ce{Nb4C3}    & 1900 & $-1.0 \times 10^{-10}$  & ~\cite{ref:lee2023investigation} \\
    \bottomrule
    \end{tabular}
    \label{tab:nonlinear_coeffs_n2}
\end{table}
 
\subsubsection{Saturable Absorption}

Saturable absorption (SA) in MXenes is an intensity-dependent nonlinear optical process in which the absorption decreases with increasing incident optical intensity. This behavior is central to passive photonic devices such as mode-locked and Q-switched lasers, as well as optical modulators and limiters operating over a broad spectral range~\cite{ref:zhao2024two, ref:shang2022nonlinear}. In MXenes, SA arises mainly from state-filling effects and nonlinear carrier dynamics. Under low excitation intensities, a large number of electronic states near the Fermi level are available for optical transitions, resulting in strong absorption. As the optical intensity increases, these states become occupied, reducing the probability of further photon absorption and thereby decreasing overall absorption. This behavior can be described by the following saturation model:
\begin{equation}
    \alpha(I) = \frac{\alpha_0}{1 + I / I_\text{sat}}
    \label{eq:SA}
\end{equation}
where $\alpha_0$ is the linear absorption coefficient and $I_\text{sat}$ is the saturation intensity, defined as the incident intensity at which the absorption reduces to half of its low-intensity value. This model captures the essential intensity-dependent response of MXenes and provides a framework for analyzing their saturable absorption behavior~\cite{ref:li2020broadband}. 

Several MXene families have been experimentally shown to exhibit strong, tunable SA under femtosecond and nanosecond excitation, with the nonlinear response depending on composition, thickness, and excitation wavelength. Ti-based MXenes, especially \ce{Ti3C2T_x}, exhibit broadband SA behavior across near- and mid-infrared wavelengths, as confirmed by open-aperture Z-scan measurements showing negative nonlinear absorption coefficients ($\beta < 0$) at \SI{800}, \SI{1064}, \SI{1550}, and \SI{1800}{nm} ~\cite{ref:jiang2018broadband}. In addition, Ti$_3$C$_2$ exhibits SA with $\beta = -5.9 \times 10^{-7}$ and $-6.8 \times 10^{-7}~\mathrm{cm/W}$ at \SI{1064}{\nano\meter} and \SI{1342}{\nano\meter}, respectively, together with a transition to reverse saturable absorption (RSA) driven by two-photon absorption (TPA) at higher excitation intensities ~\cite{ref:zhang2021third}, highlighting the tunability of the nonlinear response under varying excitation conditions. Film thickness also plays a critical role, as demonstrated by \ce{Ti3C2T_x} films of \SI{15}{nm} exhibiting $\beta = 7.13 \times10^{-7} ~\mathrm{cm/W}$, while thicker 90~\si{\nano\meter} films have $\beta = -2.69 \times 10^{-7} ~\mathrm{cm/W}$~\cite{ref:jin2024thickness}. It is worth noting that the thickness of the MXene can cause strong TPA in MXenes \cite{ref:wang2020two}. These results indicate that MXene-based SA devices can be precisely optimized for target laser systems through careful control of thickness and fabrication. Beyond Ti-based MXenes, Nb- and V-based MXenes extend the saturable absorption response toward longer wavelengths. Niobium carbides such as \ce{Nb2C} and \ce{Nb4C3} exhibit stable SA behavior in the near- and mid-infrared regions, with reported nonlinear absorption coefficients of $\beta = -8$ and $-5 \times 10^{-11} ~\mathrm{cm/W}$ at \SI{1300} and \SI{1550}{nm}, respectively, and $\beta = -2.52 \times10^{-6} ~\mathrm{cm/W}$ at \SI{1900}{nm}, respectively~\cite{ref:wang20232d, ref:lee2023investigation}. Similarly, Vanadium-based MXenes have also been shown to exhibit pronounced SA responses in the mid-infrared region, with \ce{V2C} exhibiting $\beta = -2.967 \times 10^{-4} ~\mathrm{cm/W}$ at \SI{1900}{nm}, and \ce{V4C3} showing $\beta = -0.527$ and $-1.89 \times 10^{-5} ~\mathrm{cm/W}$ at \SI{1559}{nm} and \SI{1911}{nm}, respectively~\cite{ref:lee2021investigation, ref:lee2023nonlinear}. These results highlight the strong composition-dependent tunability of nonlinear absorption in MXenes and their suitability for mid-infrared photonic applications.

\begin{figure}[htbp]
    \centering
    \includegraphics[width=1\textwidth]{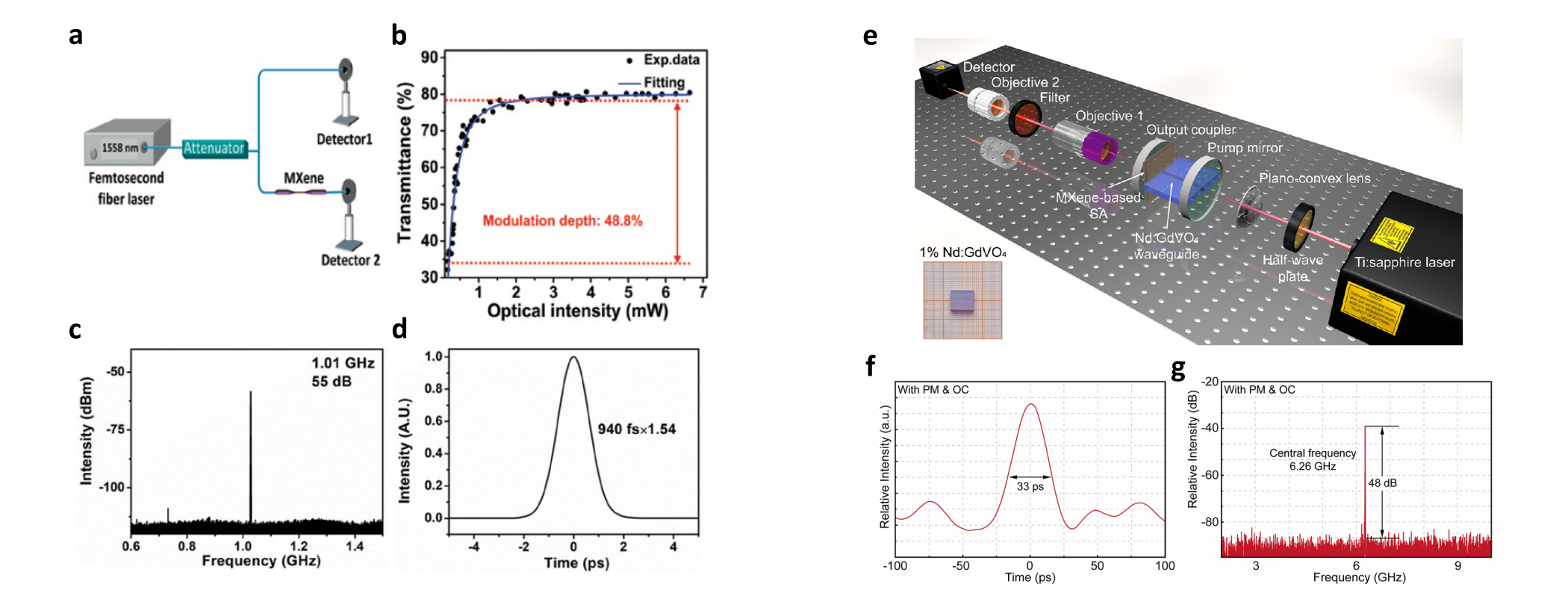}
    \caption{\rev{(a--d) Saturable absorption of \ce{V2CT_x} MXene for passively mode-locked (PML) fiber lasers, showing (a) nonlinear transmittance measurement setup, (b) transmittance versus optical intensity with fit, giving a modulation depth of 48.8\%, (c) RF spectrum of the 206th-harmonic mode-locked laser, and (d) autocorrelation trace. Adapted from~\cite{ref:huang2020highly}. (e--g) \ce{Ti3C2T_x} MXene saturable absorber in a Q-switched mode-locked (QML) waveguide laser, showing (e) the laser setup, (f) the mode-locked pulse, and (g) the RF spectrum. Adapted from~\cite{ref:li2023nanometer}.}}
    \label{fig:Fig5}
\end{figure}

SA in MXenes has been directly leveraged for practical photonic devices, including mode-locking, Q-switching, and single-frequency fiber lasers. Mode-locking enables the generation of ultra-short pulses ranging from picoseconds to femtoseconds, while Q-switching is used to produce extremely high-peak-power pulses~\cite{ref:svelto1998principles}. Traditional passive switching techniques employ SAs such as transition-metal-based materials~\cite{ref:pan2007passively}, semiconductor saturable absorber mirrors (SESAMs)~\cite{ref:huang2009high}, and carbon nanotubes~\cite{ref:zhou2009tunable, ref:liu2013carbon}. Recently, MXenes have been integrated onto waveguides to realize compact SA mirrors with broadband operation. \rev{Saturable absorption of \ce{V2CT_x} MXene was utilized for passively mode-locked (PML) fiber lasers (Fig.~\ref{fig:Fig5}a--d).} \rev{The nonlinear transmittance of the \ce{V2CT_x} saturable absorber was measured with a balanced twin-detector setup using a \SI{1558}{nm} femtosecond fiber laser (Fig.~\ref{fig:Fig5}a), and fitting the transmittance versus optical intensity gives a modulation depth of 48.8\% (Fig.~\ref{fig:Fig5}b).} \rev{By depositing \ce{V2CT_x} nanosheets on a tapered fiber in an Er-doped ring cavity, 206th-order harmonic mode-locking was obtained, with a repetition rate of 1.01 GHz and a signal-to-noise ratio of 55 dB (Fig.~\ref{fig:Fig5}c) and a pulse duration of 940 fs (Fig.~\ref{fig:Fig5}d)~\cite{ref:huang2020highly}.} \rev{In addition, MXene was used for a Q-switched mode-locked (QML) laser (Fig.~\ref{fig:Fig5}e--g).} \rev{Titanium carbide (\ce{Ti3C2T_x}) and tantalum carbide (\ce{Ta4C3T_x}) saturable absorbers, spin-coated on sapphire substrates, were combined with an \ce{Nd}:\ce{GdVO4} waveguide laser pumped by a Ti:sapphire laser (Fig.~\ref{fig:Fig5}e), enabling QML operation at a wavelength of \SI{1063}{nm}.} \rev{The \ce{Ti3C2T_x} laser delivered a mode-locked pulse duration of \SI{33}{ps} (Fig.~\ref{fig:Fig5}f) at a repetition rate of \SI{6.26}{GHz} with a signal-to-noise ratio of 48 dB (Fig.~\ref{fig:Fig5}g), while the \ce{Ta4C3T_x} laser reached \SI{30}{ps}~\cite{ref:li2023nanometer}.} Similarly, an \ce{Er}:\ce{Y2O3} ceramic channel waveguide laser was Q-switched using a \ce{Ti3C2T_x} SA mirror, achieving lasing at \SI{2710}{nm} with a pulse duration of \SI{58.1}{\nano\second}~\cite{ref:tian2020mxene}. These demonstrations highlight the integration of MXenes into compact ultrafast photonic platforms.

Under certain conditions, MXenes can exhibit reverse saturable absorption (RSA), where absorption increases with increasing optical intensity, which is relevant for optical limiting and high-power laser protection. This transition reflects the underlying carrier dynamics governing nonlinear absorption and provides a natural link to the following subsection. Table \ref{tab:nonlinear_coeffs_beta} summarizes the reported $\beta$ values for various MXenes, emphasizing the wide range of saturable absorption characteristics available through material and structural engineering.

\begin{table}[H]
    \centering
    \caption{Summary of nonlinear absorption coefficient $\beta$ values for different MXenes.}
    \rowcolors{2}{gray!8}{white}
    \begin{tabular}{cccc}
    \toprule
    \textbf{MXene} & \textbf{Wavelength (nm)} & \boldmath$\beta~(\mathrm{cm/W})$ & \textbf{Reference} \\
    \midrule
    \ce{Ti3C2T_x} & 800  & $-2.97 \times 10^{-10}$ & \cite{ref:jiang2018broadband} \\
    \ce{Ti3C2T_x} & 1064 & $-2.06 \times 10^{-10}$ & \cite{ref:jiang2018broadband} \\
    \ce{Ti3C2T_x} & 1550 & $-3.58 \times 10^{-10}$ & \cite{ref:jiang2018broadband} \\
    \ce{Ti3C2T_x} & 1800 & $-1.12 \times 10^{-10}$ & \cite{ref:jiang2018broadband} \\
    \ce{Ti3C2} & 1064 & $-5.9 \times 10^{-7}$ & \cite{ref:zhang2021third} \\
    \ce{Ti3C2} & 1342 & $-6.8 \times 10^{-7}$ & \cite{ref:zhang2021third} \\

    \ce{Ti3C2T_x} (15 nm) & 800 & $7.13 \times 10^{-7}$ & \cite{ref:jin2024thickness} \\
    \ce{Ti3C2T_x} (90 nm) & 800 & $-2.69 \times 10^{-7}$ & \cite{ref:jin2024thickness} \\
    
    \ce{Nb2C} & 1300  & $-8 \times 10^{-11}$ & \cite{ref:wang20232d} \\
    \ce{Nb2C} & 1550 & $-5 \times 10^{-11}$  & \cite{ref:wang20232d} \\
    \ce{Nb4C3} & 1900 & $-2.52 \times 10^{-6}$ & \cite{ref:lee2023investigation} \\
    
    \ce{V2C}  & 1900 & $-2.967 \times 10^{-4}$ & \cite{ref:lee2021investigation} \\
    \ce{V4C3} & 1559 & $-0.527 \times 10^{-5}$ & \cite{ref:lee2023nonlinear} \\
    \ce{V4C3} & 1911 & $-1.89 \times 10^{-5}$ & \cite{ref:lee2023nonlinear} \\
    
    \bottomrule
    \end{tabular}
    \label{tab:nonlinear_coeffs_beta}
\end{table}

\subsubsection{Nonlinear Frequency Conversion}
\begin{figure}[htbp]
    \centering
    \includegraphics[width=1\textwidth]{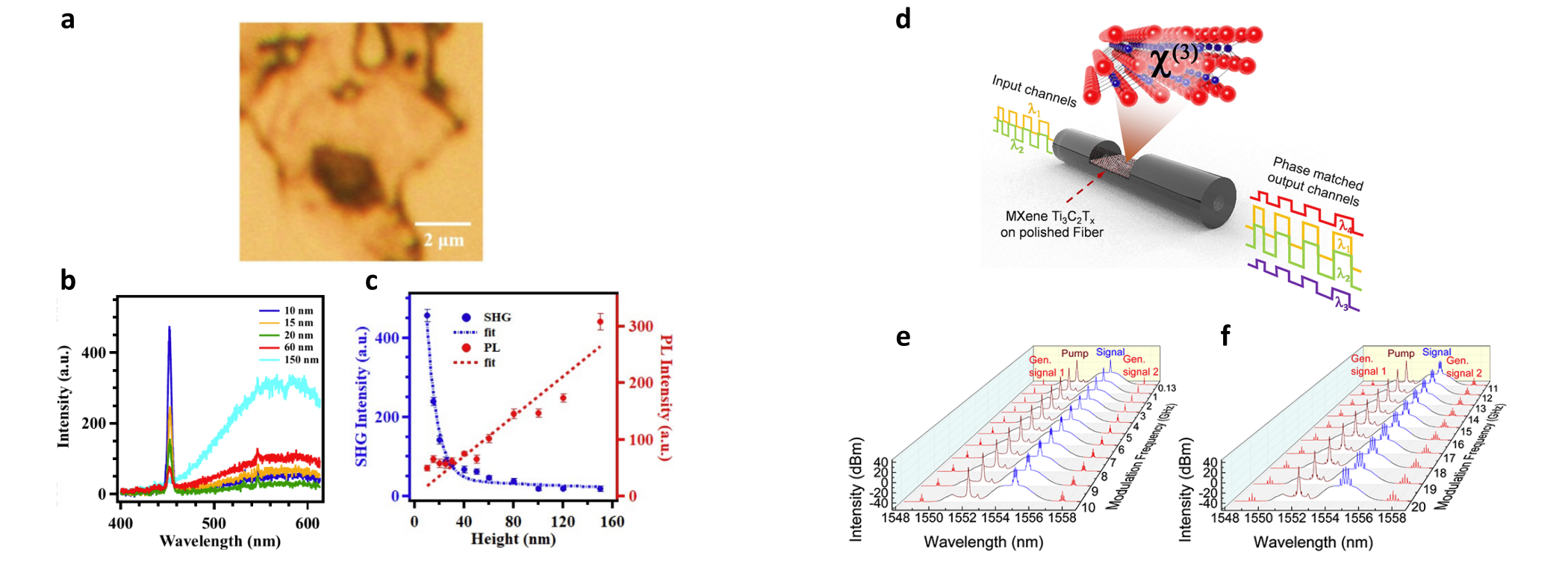}
    \caption{\rev{(a--c) \ce{Ti3C2T_x} MXene flakes on a silicon substrate for second-harmonic generation (SHG), showing (a) the optical image of a flake, (b) the emission spectra for different flake thicknesses, and (c) the SHG and two-photon fluorescence (PL) intensities versus thickness. Adapted from~\cite{ref:debow2024two}. (d--f) \ce{Ti3C2T_x} MXene on a side-polished fiber for four-wave mixing (FWM) based ultrafast optical switching, showing (d) the device schematic and (e,f) the FWM spectra for signal modulation frequencies of 0.13--10~GHz and 11--20~GHz, respectively. Adapted from~\cite{ref:uddin2024asymmetric}.}}
    \label{fig:Fig6}
\end{figure}

Although MXenes are primarily known for their strong third-order nonlinear optical response, recent studies have demonstrated their potential for harmonic generation and frequency-conversion applications. Harmonic generation remains considerably less explored than Kerr nonlinearities and saturable absorption, yet emerging experimental and theoretical investigations indicate promising opportunities for exhibiting both second- and third-order nonlinear processes in MXene-based photonic platforms. Second-harmonic generation (SHG) is generally weak in pristine MXenes because most MXene structures possess inversion symmetry, resulting in a negligible bulk $\chi^{(2)}$ response. However, SHG has been experimentally observed in \ce{Ti3C2T_x} flakes \cite{ref:debow2024two}. \rev{Flakes deposited on a silicon substrate (Fig.~\ref{fig:Fig6}a) were excited at \SI{900}{nm}, and the emission spectra for different flake thicknesses (Fig.~\ref{fig:Fig6}b) show a strong SHG peak at \SI{450}{nm} for thin flakes.} \rev{With increasing thickness, the SHG intensity decreases exponentially while the two-photon fluorescence (PL) increases linearly (Fig.~\ref{fig:Fig6}c).} Furthermore, first-principles calculations predicted that strain-induced symmetry breaking in \ce{Sc2CT2} monolayers can generate a significant second-order nonlinear response, enabling efficient SHG emission \cite{ref:he2020strain}. These findings suggest that surface engineering, strain, and interfacial effects may provide practical routes to activate $\chi^{(2)}$ nonlinearities in MXenes.

In contrast, third-harmonic generation (THG) naturally arises from the strong third-order susceptibility of MXenes and is therefore expected to dominate their harmonic-generation response. The combination of metallic conductivity, high carrier density, and ultrafast carrier dynamics makes MXenes attractive candidates for efficient THG and nonlinear frequency conversion. Although direct experimental demonstrations of THG in MXenes remain scarce, recent advances in photonic engineering provide promising routes toward enhancing third-order nonlinear interactions. In particular, Stavrou et al. demonstrated that integrating \ce{Mo2Ti2C3T_x} MXenes within slow-light woodpile photonic crystals significantly enhances effective Kerr nonlinearities through prolonged light-matter interaction times and strong electromagnetic field confinement~\cite{ref:stavrou2026slow}. Similarly, recent studies on nonlinear metasurfaces exploiting bound states in the continuum (BICs) have shown substantial enhancement of THG and degenerate four-wave mixing efficiencies owing to resonantly increased local electromagnetic fields~\cite{ref:qin2025enhanced}. These results suggest that slow-light architectures, resonant metasurfaces, and cavity-assisted platforms may offer practical pathways for realizing efficient MXene-based THG devices and nonlinear frequency converters in future integrated photonic systems.

Another important third-order nonlinear process is four-wave mixing (FWM), which enables wavelength conversion and all-optical signal processing. While FWM in MXenes remains largely unexplored, \ce{Ti3C2T_x} MXene has already been employed in an all-optical switching demonstration based on FWM interactions \cite{ref:uddin2024asymmetric}. \rev{In this device, a multilayer MXene film was deposited on the polished side of an optical fiber (Fig.~\ref{fig:Fig6}d), and the interaction of the input channels with the MXene through $\chi^{(3)}$ generates new phase-matched channels.} \rev{When the signal is modulated at frequencies from 0.13 to 10 GHz (Fig.~\ref{fig:Fig6}e) and from 11 to 20 GHz (Fig.~\ref{fig:Fig6}f), the modulation is copied to the two newly generated signals, confirming FWM operation at modulation frequencies up to 20 GHz.} This initial result suggests that MXenes may serve as efficient nonlinear media for future integrated wavelength converters and optical computing architectures.

Overall, harmonic generation in MXenes remains in its early stages compared with saturable absorption and Kerr nonlinearity. Future efforts focused on symmetry engineering, plasmonic enhancement, heterostructure design, and integrated waveguide platforms may unlock efficient SHG, THG, and FWM functionalities, establishing MXenes as versatile materials for frequency conversion and nonlinear integrated photonics.

\section{Modeling Nonlinear Effects in MXenes}
The unique electrical and optical properties of MXenes, including electrical conductivity and optical absorption, are due to their electronic band structure. Monolayer MXenes exhibit distinct band structures and electrical and optical properties compared with multilayer MXenes. Furthermore, the properties of the material can be tuned by using different terminations. Developing a modeling framework can help optimize both the material itself and the electronic and photonic devices based on MXenes.

Historically, MXenes have been used in energy storage applications~\cite{ref:thomas2023review}, and consequently, their electrical properties have been more extensively investigated than their optical properties. For example, many MXenes exhibit high carrier concentrations and high electrical conductivity, including \ce{Ti3C2}, which has an electrical conductivity of approximately \SI{2.0e4}{\siemens\per\centi\meter}~\cite{ref:mathis2021modified}. Using different terminations in \ce{Ti3C2T_x} can result in varying conductivities, ranging from \SIrange{6.0e3}{8.0e3}{\siemens\per\centi\meter}~\cite{ref:thomas2023review}. On the other hand, there are other MXenes, such as \ce{V2CT_x} and \ce{Nb2CT_x}, which exhibit lower conductivities of around \SI{1.0e3}{\siemens\per\centi\meter}~\cite{ref:matthews2021guidelines}, and \SI{5.0e2}{\siemens\per\centi\meter}~\cite{ref:zhu2024additive}, respectively.

Currently, there is much ongoing research on the use of MXenes for photonic applications, and modeling the optical properties is necessary for optimization at both the material and device levels. Calculating the band structures of MXenes can be used to model optical properties and effects, including the dielectric constant and nonlinear optical absorption~\cite{ref:minassian2025nonlinear}. Most MXenes are low-bandgap semiconductors or metallic materials~\cite{ref:tang2012mxenes}. E.g., \ce{Nb2CT_x} behaves as a dielectric while \ce{V2CT_x} exhibits metallic behavior. The monolayer and multilayer nature, including flake irregularities, different terminations, material thickness, and stoichiometry variations, make the band structure diverse and calculations more complex. There are some studies on band structure calculations for different material thicknesses~\cite{ref:saha2023surface}.

There have been many papers reporting the optical properties of MXenes recently. One of the optical properties that has been experimentally characterized in detail is the nonlinear optical absorption, including SA and RSA~\cite{ref:kim2018femtosecond, ref:jiang2018broadband}. The first experimental studies on optical nonlinearities in 2D titanium carbide and carbonitride MXenes (\ce{Ti3C2T_x} and \ce{Ti3CNT_x}, where \ce{T_x} represents functional groups like -OH and -F) confirmed their robust nonlinear optical behavior. In Ref.~\cite{ref:jhon2017metallic}, \ce{Ti3CN} is used as an SA in a mode-locked laser that produces pulses with a duration of \SI{660}{\femto\second}, a repetition rate of \SI{15.4}{\mega\hertz} at \SI{1557}{nm}. In the same work, density functional theory (DFT) was employed to demonstrate that \ce{Ti3CN} is a metallic material. Similarly, saturable absorption properties of 2D \ce{Ti3CNT_x} are characterized for different material thicknesses~\cite{ref:dong2018saturable}. The research concludes that the saturable absorption properties are dependent on material thickness, and the fundamental cause of this dependency remains unknown.

In most cases, the saturation theory of nonlinear absorption in a two-level atomic system is employed to model nonlinear optical absorption in MXenes (Eq.~\ref{eq:SA}). Fitting experimental data to determine $I_{\text{sat}}$ is used to model saturable absorption in MXenes. Nevertheless, this approach does not explain the underlying physics of nonlinear optical absorption. As reported in the literature, the $I_{\text{sat}}$ of MXenes depends on the specific material. The importance of the band structure is further underscored by the observed discrepancies in saturation intensity across experimental studies~\cite{ref:minassian2025nonlinear}. $I_{\text{sat}}$ of \SI{1.29}{\mega\watt\per\centi\meter\squared} at \SI{1000}{nm} and \SI{1.02}{\mega\watt\per\centi\meter\squared} at \SI{1500}{nm} is reported in Ref.~\cite{ref:shang2022nonlinear}; \SI{10.68}{\mega\watt\per\centi\meter\squared} at \SI{1550}{nm} and \SI{651.23}{\mega\watt\per\centi\meter\squared} at \SI{2000}{nm} are reported in Ref.~\cite{ref:wang2020mxene}. Such wide variations are reasonably associated with the unique peculiarities of the MXene band structure. The model developed in \cite{ref:minassian2025nonlinear}, based on the electronic band structure, can describe nonlinear optical absorption. It was validated with \ce{Ti3C2T_x}--PVA experimental data at \SI{1064}{nm}~\cite{ref:rosol2024short}, \SI{1090}{nm}~\cite{ref:wang2024high}, and \SI{1565}{nm}~\cite{ref:liang2023passively}.

Modeling nonlinear optical effects in MXenes is crucial to advancing their integration into photonic devices and to understanding the underlying physics. While their electrical properties have been extensively studied due to their origins in energy-related applications, the optical and, particularly, the nonlinear optical behavior of MXenes is a rapidly emerging area of research. The continued development of such modeling can be extended to other nonlinear effects observed in materials, including second- and third-harmonic generation and four-wave mixing. \\

\section{Applications}
\subsection{Sensing systems}
\begin{figure}[htbp]
    \centering
    \includegraphics[width=\textwidth]{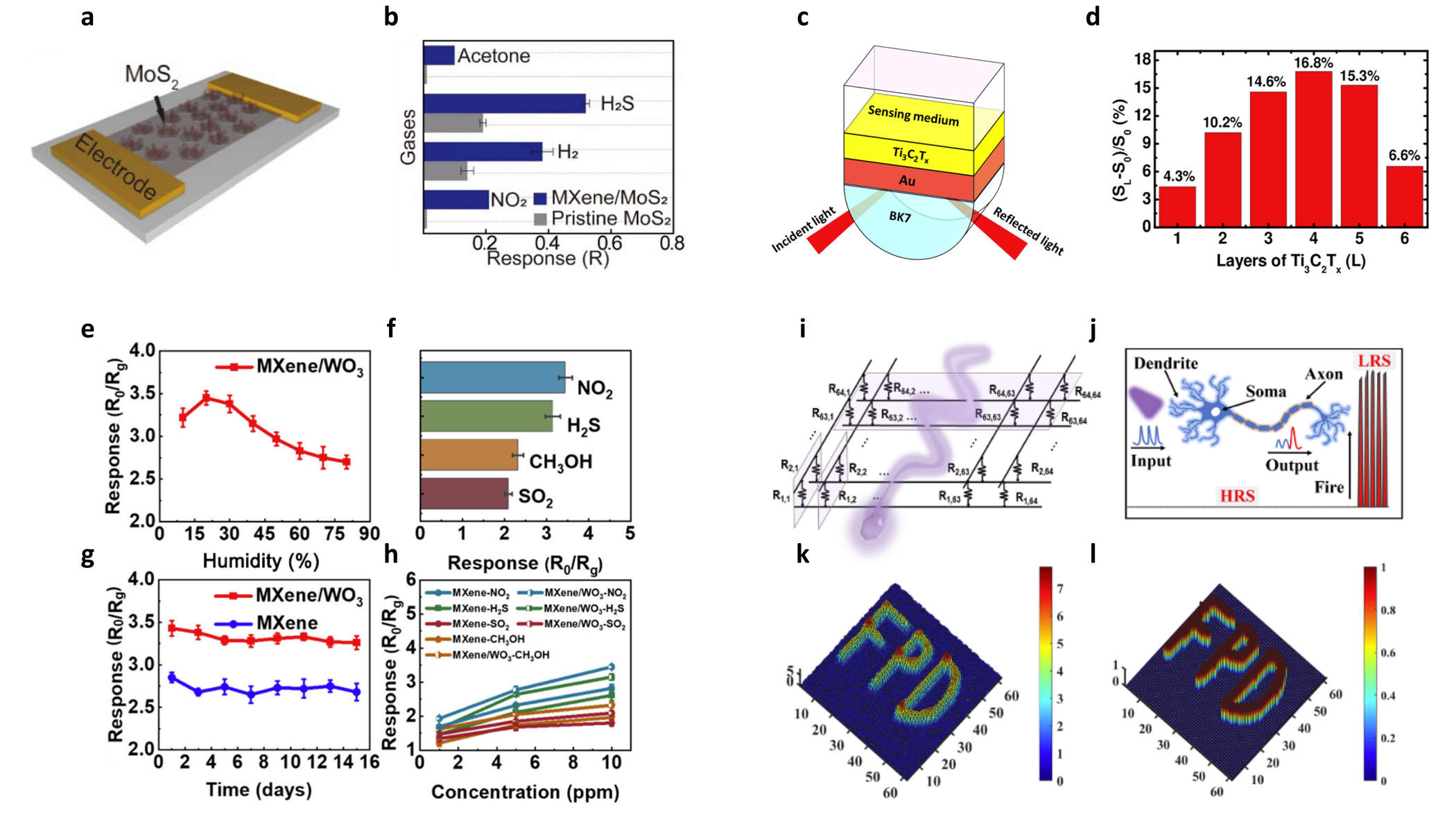}
    \caption{\rev{MXene-based gas sensing paradigms spanning electrical, optical, and neuromorphic platforms. (a,b) Chemiresistive \ce{Ti3C2T_x}/\ce{MoS2} gas sensor, showing (a) the device schematic and (b) the responses of \ce{Ti3C2T_x}/\ce{MoS2} and pristine \ce{MoS2} to different gases at \SI{100}{ppm}. Adapted from~\cite{ref:shin2024high}. (c,d) SPR sensor with a few-layer \ce{Ti3C2T_x} coating, showing (c) the device schematic and (d) the sensitivity enhancement versus the number of \ce{Ti3C2T_x} layers. Adapted from~\cite{ref:wu2018few}. (e--h) \ce{V2CT_x} MXene and MXene/\ce{WO3} gas sensors, showing the response (e) versus relative humidity, (f) to different gases, (g) over 15 days, and (h) versus gas concentration. Adapted from~\cite{ref:guo2024mxene}. (i--l) Oxidized-MXene optoelectronic neuron, showing (i) the schematic of the $64\times64$ array, (j) the resistance-state switching under photoelectric stimulation, (k) the input light trajectory, and (l) the output trajectory processed by the array. Adapted from~\cite{ref:zeng2023mxene}.}}
    \label{fig:mxene_gas_sensing}
\end{figure}

MXenes have recently emerged as promising materials for optical gas sensing because of their unique structural and optical properties. Their 2D morphology, large surface area, and adjustable surface terminations, which enable strong adsorption and charge transfer with gas molecules and make them suitable for detecting toxic and volatile gases at room temperature, make them ideal for real-time monitoring in environmental, industrial, and biomedical contexts ~\cite{ref:qian2025recent, ref:lu20212d}. These properties have led to extensive exploration of MXene-based sensing systems across electrical, optical, and hybrid platforms, as illustrated in Fig.~\ref{fig:mxene_gas_sensing}. However, within the context of nonlinear photonics, the most relevant sensing mechanisms are those that directly couple gas adsorption to optical and nonlinear optical responses.

Conventional MXene-based gas sensors are primarily based on chemiresistive mechanisms, where adsorption-induced charge transfer modulates electrical conductivity~\cite{ref:kim2018metallic, ref:john2023gas}. For example, \ce{Ti3C2T_x}/\ce{MoS2} heterostructures operating at \SI{250}{\degreeCelsius} have demonstrated enhanced gas sensing performance through the formation of heterojunctions, which facilitate efficient charge transfer and increase the density of active adsorption sites. \rev{In this device, flower-like \ce{MoS2} grown by chemical vapor deposition on a Langmuir--Blodgett \ce{Ti3C2T_x} film lies between two electrodes (Fig.~\ref{fig:mxene_gas_sensing}a).} \rev{Among the gases tested, the response is highest for \ce{H2S}, reaching approximately 0.5 at \SI{100}{ppm} and clearly exceeding that of pristine \ce{MoS2} (Fig.~\ref{fig:mxene_gas_sensing}b).} \rev{Compared with pristine MXene, the heterostructure improves the \ce{H2S} response about fivefold, and the calculated detection limit reaches approximately \SI{0.87}{ppb}~\cite{ref:shin2024high},} highlighting how interface engineering can significantly enhance gas sensing performance through improved charge transfer and carrier modulation. While these approaches rely on electrical readout, optical sensing platforms enable direct transduction of molecular adsorption into measurable optical signals. \rev{MXenes have been integrated into surface plasmon resonance (SPR) configurations, in which a few-layer \ce{Ti3C2T_x} film coats the gold layer of a prism-coupled sensor (Fig.~\ref{fig:mxene_gas_sensing}c).} \rev{At a wavelength of \SI{633}{nm}, numerical modelling predicts that this coating increases the sensitivity from approximately \SI{137}{\degree\per\RIU} for a conventional Au-based sensor to about \SI{160}{\degree\per\RIU}, corresponding to an enhancement of \SI{\sim 16.8}{\percent}~\cite{ref:wu2018few}.} \rev{The enhancement peaks at four \ce{Ti3C2T_x} layers and decreases for thicker coatings (Fig.~\ref{fig:mxene_gas_sensing}d), and it is attributed to the absorption of the MXene layer,} providing a foundation for highly sensitive refractive index detection and optical gas sensing platforms.

Recent advances have further extended MXene-based sensing toward high-Q resonant photonic architectures. In particular, MXene-integrated metasurfaces supporting quasi-bound states in the continuum (quasi-BIC) have emerged as highly sensitive optical sensing platforms. These systems exploit symmetry-protected resonances to achieve strong electromagnetic field confinement and narrow spectral linewidths, significantly enhancing light-matter interaction. For example, MXene-coated dielectric metasurfaces operating near the methane absorption wavelength (\SI{\sim 1650}{nm}) have been reported to exhibit absorption exceeding \SI{90}{\percent}, quality factors greater than 150, and refractive index sensitivities on the order of \SI{\sim 170}{\nano\meter\per\RIU}. In addition, measurable wavelength shifts of approximately \SI{\sim 0.8}{\nano\meter\per\percent} gas concentration enable precise detection through resonance tracking~\cite{ref:sharma2025bound}. Such high-Q resonant platforms provide a powerful route for enhancing sensing performance, demonstrating the potential of combining MXenes with engineered photonic structures for compact, high-sensitivity, and wavelength-selective gas sensing. Furthermore, MXene-based sensing platforms have recently been extended toward intelligent and neuromorphic gas sensing systems. Sensor arrays composed of \ce{V2CT_x} MXene and MXene/\ce{WO3} nanocomposites have been integrated with spiking neural networks (SNNs) to enable the identification of four toxic gases~\cite{ref:guo2024mxene}. \rev{Adding \ce{WO3} nanoparticles raises the response to \SI{10}{ppm} \ce{NO2} from 2.82 to 3.45, and the sensor responds most strongly to \ce{NO2} among \ce{CH3OH}, \ce{H2S}, \ce{NO2} and \ce{SO2} (Fig.~\ref{fig:mxene_gas_sensing}f).} \rev{The response first rises and then falls with relative humidity, from 3.45 at \SI{20}{\percent} to 2.70 at \SI{80}{\percent} (Fig.~\ref{fig:mxene_gas_sensing}e).} \rev{It remains almost unchanged over 15 days (Fig.~\ref{fig:mxene_gas_sensing}g) and increases with gas concentration up to \SI{10}{ppm} (Fig.~\ref{fig:mxene_gas_sensing}h).} \rev{The SNN encodes the temporal response curves into spike sequences and classifies the four gases with an accuracy of \SI{95.83}{\percent}~\cite{ref:guo2024mxene}.} This combination of sensing and neuromorphic computation highlights the potential of MXene-based systems for intelligent and adaptive sensing platforms beyond conventional single-analyte detection.

Beyond gas recognition, MXene-based neuromorphic sensing architectures have recently been extended toward multifunctional optoelectronic systems capable of simultaneously performing sensing, signal preprocessing, and adaptive information processing. In particular, Zeng \textit{et al.} demonstrated an artificial optoelectronic neuron based on an Ag/MXene/SiO$_2$/Si memristive architecture, where an oxidized MXene layer enabled efficient photoelectric integration under synergistic optical and electrical stimulation~\cite{ref:zeng2023mxene}. \rev{A $64\times64$ array of these devices is used for sensing and processing light trajectories (Fig.~\ref{fig:mxene_gas_sensing}i).} \rev{Each device mimics a biological neuron by firing, that is, by switching from a high-resistance state to a low-resistance state once the integrated input exceeds a threshold (Fig.~\ref{fig:mxene_gas_sensing}j).} \rev{The array converts a moving light trajectory (Fig.~\ref{fig:mxene_gas_sensing}k) into resistance states and sharpens it, so that the output trajectory (Fig.~\ref{fig:mxene_gas_sensing}l) has higher contrast and fewer redundant signals at its edges,} illustrating the potential of MXene-based neuromorphic devices for intelligent sensing platforms that integrate perception and computation within a unified hardware framework.

Although most MXene-based gas sensors rely on electrical or linear optical transduction mechanisms, the underlying adsorption-induced changes in carrier density and dielectric properties are inherently linked to the nonlinear optical response of MXenes. In particular, variations in carrier concentration and interfacial interactions can influence the nonlinear refractive index ($n_2$) and nonlinear absorption coefficient ($\beta$), suggesting that gas adsorption may dynamically modulate the third-order susceptibility ($\chi^{(3)}$). This provides a potential pathway toward nonlinear photonic gas sensing, where molecular interactions are transduced through intensity-dependent optical responses, enabling ultrafast and all-optical detection schemes.

\subsection{Photonic neural networks}
MXenes have recently emerged as promising materials for photonic neural networks (PNNs) due to their strong and ultrafast NLO response. Their large third-order susceptibility ($\chi^{(3)}$), broadband optical absorption, and tunable carrier dynamics make them attractive candidates for implementing compact nonlinear elements in photonic systems. In contrast to conventional electronic neural networks, which are limited by bandwidth, latency, and energy consumption, PNNs exploit the inherent parallelism and high speed of light to enable ultrafast and energy-efficient information processing~\cite{ref:zou2008overview}. However, the realization of fully optical neural networks critically depends on the availability of materials capable of providing strong, fast, and reconfigurable nonlinear activation functions. As shown in prior work~\cite{ref:shastri2021photonics}, the lack of such materials remains a key bottleneck in scalable photonic neuromorphic hardware. MXenes address several of these requirements, positioning them as promising building blocks for nonlinear photonic computing.

\begin{figure}[htbp]
    \centering
    \includegraphics[width=\textwidth]{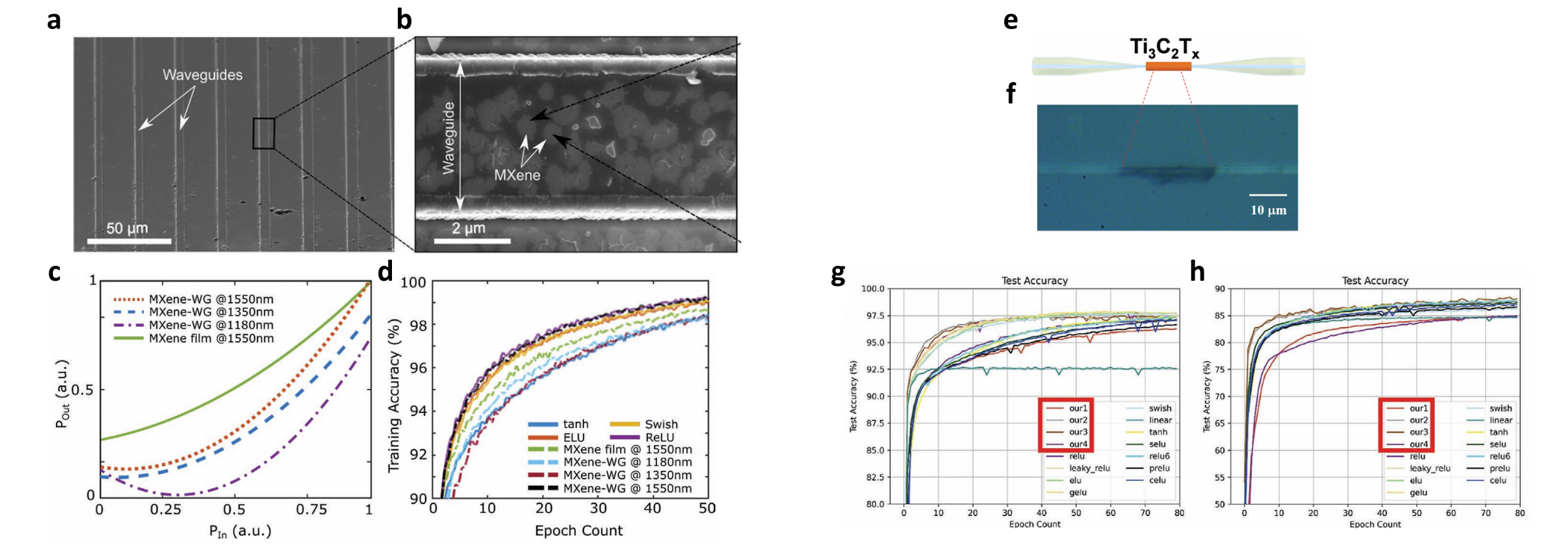}
    \caption{\rev{MXene-enabled nonlinear photonic devices for neural network applications. (a--d) \ce{Ti3C2T_x} MXene flakes on a silicon rib waveguide for all-optical nonlinear activation, showing (a) the top-view SEM image of the rib waveguides, (b) the zoomed SEM image of a waveguide covered with MXene flakes, (c) the activation functions at different operating wavelengths for the MXene-covered waveguide and an MXene thin film, and (d) the training accuracy versus epoch count for MNIST classification compared with software activation functions. Adapted from~\cite{ref:hazan2023mxene}. (e--h) \ce{Ti3C2T_x}-coated microfiber nonlinear activator, showing (e) the device schematic, (f) the optical microscope image of the coating, and the test accuracy versus epoch count for (g) MNIST and (h) Fashion-MNIST classification using the four measured activation functions (labeled our1 to our4 in the original paper) and software activation functions. Adapted from~\cite{ref:yang2022mxene}.}}
    \label{fig:MXene_NN_fig}
\end{figure}

Recent studies have demonstrated that MXenes can serve as effective nonlinear activation elements in photonic neural networks. In particular, \ce{Ti3C2T_x} MXene flakes have been deposited on a silicon rib waveguide to realize an all-optical nonlinear activation function directly on a chip~\cite{ref:hazan2023mxene}. \rev{SEM images show the rib waveguides and the MXene flakes covering one of them (Fig.~\ref{fig:MXene_NN_fig}a,b), and the light interacts with the flakes through its evanescent field.} \rev{The resulting intensity-dependent transmission, which the authors attribute to the saturable absorption of MXene, depends on the operating wavelength, so the shape of the activation function can be tuned between \SI{1180}{nm} and \SI{1550}{nm} without changing the device (Fig.~\ref{fig:MXene_NN_fig}c).} \rev{In a neural-network emulation that uses these measured activation functions, the classification accuracy on the MNIST handwritten digit dataset lies between \SI{97.9}{\percent} and \SI{99.1}{\percent}, comparable to that of software activation functions (Fig.~\ref{fig:MXene_NN_fig}d)~\cite{ref:hazan2023mxene}.} \rev{Complementary approaches using MXene-coated optical microfibers have further demonstrated broadband and ultrafast nonlinear activation, in which a \ce{Ti3C2T_x} coating on a microfiber drawn from a single-mode fiber interacts strongly with the evanescent field (Fig.~\ref{fig:MXene_NN_fig}e,f)~\cite{ref:yang2022mxene}.} \rev{Depending on the length of MXene covering the microfiber, the transmission is governed by SA or RSA, which yields four activators with different activation functions.} \rev{Using these measured activation functions in three-layer neural-network simulations, the SA activator with the largest modulation depth reaches \SI{97.54}{\percent} on MNIST and \SI{88.01}{\percent} on Fashion-MNIST, comparable to ReLU (Fig.~\ref{fig:MXene_NN_fig}g,h)~\cite{ref:yang2022mxene}.} \rev{The carrier relaxation of the \ce{Ti3C2T_x} MXene is completed within \SI{5.22}{ps}, which corresponds to a calculated operation frequency exceeding \SI{190}{\giga\hertz}~\cite{ref:yang2022mxene}.} Together, these results establish MXenes as efficient and ultrafast nonlinear activation media for photonic neural networks. Recently, MXenes have been extended beyond nonlinear activation elements toward fully programmable optical computing platforms. Ge et al. demonstrated an electrically reconfigurable all-optical logic processing unit (LPU) based on the Kerr nonlinearity of high-entropy \ce{TiVCrMoC3T_x} MXene combined with electrochemically tunable surface terminations. The proposed architecture enabled dynamic switching among seven fundamental Boolean operations, including AND, OR, NOT, NOR, NAND, XOR, and XNOR, within a single optical configuration. By integrating these reconfigurable logic units into a three-layer optical differentiable logic gate network (DLGN), the authors further realized an end-to-end in-optics handwritten digit recognition system, achieving a classification accuracy of $97.7\%$ on the MNIST dataset. This work represents an important step toward scalable, energy-efficient, and interpretable neuromorphic photonic hardware based on MXene nonlinearities~\cite{ref:ge2026all}. Beyond programmable optical logic, MXenes have also demonstrated potential for scalable neuromorphic hardware implementations. Huang \textit{et al.} reported a flexible, transparent, and wafer-scale \ce{TiO_x/Ti3C2T_x} artificial synapse array capable of emulating synaptic weight modulation and pattern-recognition tasks, achieving a classification accuracy of $96.44\%$ on the MNIST dataset, thereby highlighting the potential of MXenes for large-area neuromorphic hardware implementations~\cite{ref:huang2023flexible}. Furthermore, oxidized \ce{Ti3C2T_x}-based memristive devices have been shown to exhibit biological synaptic plasticity characteristics, including paired-pulse facilitation (PPF), long-term potentiation (LTP), and long-term depression (LTD), highlighting the broader potential of MXenes for adaptive neuromorphic functionalities and suggesting opportunities for future integration with photonic computing architectures~\cite{ref:feng2023novel, ref:wang2024dual}.

The nonlinear optical response of MXenes can be significantly enhanced through structural design and plasmonic engineering. For example, MXene-based metasurface resonators have demonstrated ultrahigh nonlinear optical responses based on acoustic MXene plasmons (AMPs) in hybrid \ce{Ti3C2T_x}-dielectric-metal nanostructures~\cite{ref:park2024ultrahigh}. In such configurations, strong electromagnetic field confinement within a nanoscale spacer enables the formation of highly localized plasmon modes in the short-wave infrared (SWIR) region (\SIrange{1.5}{6}{\micro\meter}), with plasmon wavelengths reduced to as small as approximately $\lambda_0/112$. This extreme confinement results in dramatically enhanced light--matter interaction, giving rise to nonlinear absorption coefficients on the order of $\beta \approx$ \SI{1.37e-2}{\meter\per\watt} at \SI{1560}{\nano\meter}, corresponding to an enhancement of nearly three orders of magnitude compared to conventional 2D materials such as graphene and black phosphorus. In addition, strong third-harmonic generation (THG) and Kerr-type nonlinearities have been observed, confirming the exceptionally large third-order susceptibility of the system. This plasmon-enhanced mechanism provides a powerful route toward compact and highly efficient nonlinear elements for dense photonic integration. In addition, the nonlinear response of MXenes can be actively tuned to achieve reconfigurable activation behavior. By controlling film thickness and excitation wavelength, MXene thin films exhibit a transition between SA and RSA. Specifically, RSA behavior is observed near \SI{800}{\nano\meter}, while SA dominates at telecommunications wavelengths around \SI{1550}{\nano\meter} when integrated onto silicon nitride waveguides~\cite{ref:jin2024thickness}. This tunability enables the design of adaptable nonlinear transfer functions for photonic neural networks.

At the material level, MXenes also exhibit ultrafast carrier dynamics and strong Kerr nonlinearity, further supporting their application in high-speed photonic computing. Studies on various MXene compositions, including \ce{Ta4C3T_x} and \ce{Mo2Ti2C3T_x}, have revealed sub-picosecond carrier relaxation times, picosecond recombination processes, and exceptionally large third-order nonlinear susceptibilities~\cite{ref:stavrou2025emerging}. These properties enable ultrafast switching and suggest potential applications in optical memory and neuromorphic photonic circuits. In addition to feedforward photonic neural networks, MXenes may also enable alternative photonic computing paradigms such as reservoir computing. In such systems, a nonlinear dynamical element with inherent memory transforms temporal input signals into high-dimensional representations, enabling efficient processing of time-dependent data. All-optical reservoir computing architectures rely on nonlinear optical components, such as saturable absorbers, to perform this transformation. Early demonstrations have shown that optical nonlinearities based on saturation effects can serve as the core mechanism for such systems~\cite{ref:duport2012all}. More recently, scalable photonic reservoir computers employing nonlinear amplifying loop mirrors (NALMs), integrated nonlinear waveguides, and wavelength-division multiplexing have demonstrated parallel processing capabilities with computational throughputs reaching 20~TOPS and energy efficiencies as low as 4.4~fJ per operation, highlighting the rapid progress toward practical neuromorphic photonic processors~\cite{ref:aadhi2025scalable}. Given their strong, ultrafast, and tunable nonlinear optical response, MXenes represent promising candidates for implementing compact and high-speed photonic reservoir computing platforms.

Overall, these studies highlight the versatility of MXenes as nonlinear photonic materials for artificial intelligence applications. Through integration into waveguides, metasurfaces, and hybrid optical platforms, MXenes enable scalable, low-power, and ultrafast photonic neural systems, positioning them as a key material platform for next-generation optical computing technologies.

\section{Future Directions}
The field of nonlinear photonics has witnessed a paradigm shift with the emergence of MXenes, offering an exceptional combination of broadband absorption, ultrafast carrier dynamics, and high third-order nonlinearities. Yet, their integration into real-world photonic systems remains at a formative stage. Looking ahead, several promising directions can guide future research and development, aimed at harnessing the full potential of MXenes for next-generation optical technologies.

A primary research priority is the enhancement of nonlinear optical performance through advanced material engineering. Tailoring surface terminations-such as replacing common hydroxyl or fluoride groups with halogens, chalcogens, or organic ligands-can significantly alter electronic structure and improve nonlinear susceptibilities. Furthermore, constructing MXene-based heterostructures with other 2D materials like \ce{MoS2}, graphene, or black phosphorus offers a pathway toward synergistic enhancement of optical nonlinearities and spectral tunability. Defect engineering and doping strategies also hold great promise in modulating mid-gap states and boosting carrier mobility, both of which are critical for processes like two-photon absorption and the optical Kerr effect.

Expanding the application landscape of MXenes represents another critical direction. Thus far, research has predominantly focused on optical limiters and saturable absorbers for mode-locked lasers. However, MXenes’ unique combination of optical conductivity and fast response times positions them as strong contenders for neuromorphic photonics, particularly in reservoir computing and photonic neural networks. In the realm of quantum photonics, their strong third-order nonlinearities could be exploited for entangled photon pair generation via four-wave mixing, opening avenues for on-chip quantum light sources. Moreover, incorporating MXenes into nonlinear metasurfaces could enable actively tunable, ultrathin platforms for beam shaping, harmonic generation, and polarization control, bringing multifunctionality to compact optical systems.

Advancements in experimental characterization and theoretical modeling are crucial for a deeper understanding and performance optimization. Ultrafast techniques such as transient absorption spectroscopy and two-dimensional Fourier transform spectroscopy can help decipher carrier dynamics and coherent interactions within MXenes. In situ measurements under variable environmental conditions-such as temperature, humidity, or electromagnetic interference-will be necessary to assess stability and viability in real-world settings. From a computational standpoint, ab initio simulations, including time-dependent density functional theory (TD-DFT), can predict how changes in structure, defects, or surface chemistry influence nonlinear optical behavior, guiding rational material design.

Despite these promising traits, the scalability and sustainability of MXene synthesis remain challenges that must be addressed. Conventional methods relying on hazardous hydrofluoric acid (HF) pose environmental and safety concerns. Thus, future efforts should prioritize HF-free or electrochemical etching routes that maintain control over surface terminations while ensuring reproducibility. Direct growth strategies, such as chemical vapor deposition (CVD), offer an alternative by enabling substrate-integrated synthesis without the intermediate MAX phase. Establishing standardized fabrication protocols will be essential for both fundamental research and industrial translation.

Finally, realizing the full potential of MXenes in integrated photonics will require robust approaches to device integration. Techniques for hybrid integration with silicon, lithium niobate, or polymeric photonic platforms will enable the fabrication of high-speed, tunable devices such as all-optical modulators and filters. Printing and transfer methods must be optimized to deposit MXenes uniformly over large areas, particularly on flexible or curved substrates. Additionally, embedding MXene films into femtosecond laser-written waveguide circuits could lead to fully integrated nonlinear photonic chips for both classical and quantum applications.

In summary, MXenes offer a fertile ground for interdisciplinary innovation at the intersection of materials science, photonics, and nanotechnology. With strategic advances in material synthesis, nonlinear characterization, application development, and scalable integration, MXenes are poised to play a transformative role in the future of nonlinear optics. From ultrafast information processing to quantum communication and neuromorphic computing, their potential impact spans a wide array of technologies. The coming years promise to be pivotal in shaping MXenes from emerging materials into enabling platforms for next-generation photonic systems.

\section{Acknowledgements}
This work was supported by the Binational Science Foundation (BSF) grant no. 2022014.

\bibliographystyle{MSP}
\bibliography{references}

@book{ref:duarte2017tunable,
  title={Tunable laser optics},
  author={Duarte, Frank J},
  year={2017},
  publisher={CRC Press}
}

@article{ref:bogaerts2020programmable,
  title={Programmable photonic circuits},
  author={Bogaerts, Wim and P{\'e}rez, Daniel and Capmany, Jos{\'e} and Miller, David AB and Poon, Joyce and Englund, Dirk and Morichetti, Francesco and Melloni, Andrea},
  journal={Nature},
  volume={586},
  number={7828},
  pages={207--216},
  year={2020},
  publisher={Nature Publishing Group UK London}
}

@book{ref:agarwal2012quantum,
  title={Quantum optics},
  author={Agarwal, Girish S},
  year={2012},
  publisher={Cambridge University Press}
}

@book{ref:agrawal2012fiber,
  title={Fiber-optic communication systems},
  author={Agrawal, Govind P},
  year={2012},
  publisher={John Wiley \& Sons}
}

@article{ref:kumbhakar2021advance,
  title={Advance optical properties and emerging applications of 2D materials},
  author={Kumbhakar, Partha and Chowde Gowda, Chinmayee and Tiwary, Chandra Sekhar},
  journal={Frontiers in Materials},
  volume={8},
  pages={721514},
  year={2021},
  publisher={Frontiers Media SA}
}

@article{ref:hamed2020graphene,
  title={Graphene-based frequency-conversion mixers for high-frequency applications},
  author={Hamed, Ahmed and Saeed, Mohamed and Negra, Renato},
  journal={IEEE Transactions on Microwave Theory and Techniques},
  volume={68},
  number={6},
  pages={2090--2096},
  year={2020},
  publisher={IEEE}
}

@article{ref:karimi2023plasmon,
  title={Plasmon-enhanced optical nonlinearity in graphene nanomeshes},
  author={Karimi, F and Mitra, S and Soleimanikahnoj, S and Knezevic, I},
  journal={Physical Review B},
  volume={108},
  number={3},
  pages={035414},
  year={2023},
  publisher={APS}
}

@article{ref:barsoum2011elastic,
  title={Elastic and mechanical properties of the MAX phases},
  author={Barsoum, Michel W and Radovic, Miladin},
  journal={Annual review of materials research},
  volume={41},
  number={1},
  pages={195--227},
  year={2011},
  publisher={Annual Reviews}
}

@article{ref:radovic2013max,
  title={MAX phases: bridging the gap between metals and ceramics},
  author={Radovic, Miladin and Barsoum, Michel W},
  journal={American Ceramics Society Bulletin},
  volume={92},
  number={3},
  pages={20--27},
  year={2013},
  publisher={American Ceramic Society, Inc., 735 Ceramic Place, Westerville OH 43081~…}
}

@article{ref:barsoum2000mn,
  title={The MN+ 1AXN phases: A new class of solids: Thermodynamically stable nanolaminates},
  author={Barsoum, Michel W},
  journal={Progress in solid state chemistry},
  volume={28},
  number={1-4},
  pages={201--281},
  year={2000},
  publisher={Elsevier}
}

@article{ref:sokol2019chemical,
  title={On the chemical diversity of the MAX phases},
  author={Sokol, Maxim and Natu, Varun and Kota, Sankalp and Barsoum, Michel W},
  journal={Trends in Chemistry},
  volume={1},
  number={2},
  pages={210--223},
  year={2019},
  publisher={Elsevier}
}

@article{ref:tallman2013critical,
  title={A critical review of the oxidation of Ti2AlC, Ti3AlC2 and Cr2AlC in air},
  author={Tallman, Darin J and Anasori, Babak and Barsoum, Michel W},
  journal={Materials Research Letters},
  volume={1},
  number={3},
  pages={115--125},
  year={2013},
  publisher={Taylor \& Francis}
}

@article{ref:lim2022fundamentals,
  title={Fundamentals of MXene synthesis},
  author={Lim, Kang Rui Garrick and Shekhirev, Mikhail and Wyatt, Brian C and Anasori, Babak and Gogotsi, Yury and Seh, Zhi Wei},
  journal={Nature Synthesis},
  volume={1},
  number={8},
  pages={601--614},
  year={2022},
  publisher={Nature Publishing Group UK London}
}

@article{ref:maleski2017dispersions,
  title={Dispersions of two-dimensional titanium carbide MXene in organic solvents},
  author={Maleski, Kathleen and Mochalin, Vadym N and Gogotsi, Yury},
  journal={Chemistry of Materials},
  volume={29},
  number={4},
  pages={1632--1640},
  year={2017},
  publisher={ACS Publications}
}

@article{ref:khaledialidusti2021high,
  title={High-throughput computational discovery of ternary-layered MAX phases and prediction of their exfoliation for formation of 2D MXenes},
  author={Khaledialidusti, Rasoul and Khazaei, Mohammad and Khazaei, Somayeh and Ohno, Kaoru},
  journal={Nanoscale},
  volume={13},
  number={15},
  pages={7294--7307},
  year={2021},
  publisher={Royal Society of Chemistry}
}

@article{ref:zhang2024delamination,
  title={Delamination of chlorine-terminated MXene produced using molten salt etching},
  author={Zhang, Teng and Shevchuk, Kateryna and Wang, Ruocun John and Kim, Hyunho and Hourani, Jamal and Gogotsi, Yury},
  journal={Chemistry of materials},
  volume={36},
  number={4},
  pages={1998--2006},
  year={2024},
  publisher={ACS Publications}
}

@article{ref:arole2021water,
  title={Water-dispersible Ti3C2Tz MXene nanosheets by molten salt etching},
  author={Arole, Kailash and Blivin, Jackson W and Saha, Sanjit and Holta, Dustin E and Zhao, Xiaofei and Sarmah, Anubhav and Cao, Huaixuan and Radovic, Miladin and Lutkenhaus, Jodie L and Green, Micah J},
  journal={Iscience},
  volume={24},
  number={12},
  year={2021},
  publisher={Elsevier}
}

@article{ref:li2020general,
  title={A general Lewis acidic etching route for preparing MXenes with enhanced electrochemical performance in non-aqueous electrolyte},
  author={Li, Youbing and Shao, Hui and Lin, Zifeng and Lu, Jun and Liu, Liyuan and Duployer, Benjamin and Persson, Per O{\AA} and Eklund, Per and Hultman, Lars and Li, Mian and others},
  journal={Nature materials},
  volume={19},
  number={8},
  pages={894--899},
  year={2020},
  publisher={Nature Publishing Group UK London}
}

@article{ref:li2019element,
  title={Element replacement approach by reaction with Lewis acidic molten salts to synthesize nanolaminated MAX phases and MXenes},
  author={Li, Mian and Lu, Jun and Luo, Kan and Li, Youbing and Chang, Keke and Chen, Ke and Zhou, Jie and Rosen, Johanna and Hultman, Lars and Eklund, Per and others},
  journal={Journal of the American Chemical Society},
  volume={141},
  number={11},
  pages={4730--4737},
  year={2019},
  publisher={ACS Publications}
}

@article{ref:kamysbayev2020covalent,
  title={Covalent surface modifications and superconductivity of two-dimensional metal carbide MXenes},
  author={Kamysbayev, Vladislav and Filatov, Alexander S and Hu, Huicheng and Rui, Xue and Lagunas, Francisco and Wang, Di and Klie, Robert F and Talapin, Dmitri V},
  journal={Science},
  volume={369},
  number={6506},
  pages={979--983},
  year={2020},
  publisher={American Association for the Advancement of Science}
}

@article{ref:liu2023bacterial,
  title={Bacterial-mediated tumor therapy: old treatment in a new context},
  author={Liu, Yao and Niu, Lili and Li, Nannan and Wang, Yang and Liu, Mingyang and Su, Xiaomin and Bao, Xuhui and Yin, Bo and Shen, Shun},
  journal={Advanced Science},
  volume={10},
  number={12},
  pages={2205641},
  year={2023},
  publisher={Wiley Online Library}
}

@article{ref:ding2023chemical,
  title={Chemical scissor--mediated structural editing of layered transition metal carbides},
  author={Ding, Haoming and Li, Youbing and Li, Mian and Chen, Ke and Liang, Kun and Chen, Guoxin and Lu, Jun and Palisaitis, Justinas and Persson, Per O{\AA} and Eklund, Per and others},
  journal={Science},
  volume={379},
  number={6637},
  pages={1130--1135},
  year={2023},
  publisher={American Association for the Advancement of Science}
}

@article{ref:wang2023direct,
  title={Direct synthesis and chemical vapor deposition of 2D carbide and nitride MXenes},
  author={Wang, Di and Zhou, Chenkun and Filatov, Alexander S and Cho, Wooje and Lagunas, Francisco and Wang, Mingzhan and Vaikuntanathan, Suriyanarayanan and Liu, Chong and Klie, Robert F and Talapin, Dmitri V},
  journal={Science},
  volume={379},
  number={6638},
  pages={1242--1247},
  year={2023},
  publisher={American Association for the Advancement of Science}
}

@article{ref:xu2015large,
  title={Large-area high-quality 2D ultrathin Mo2C superconducting crystals},
  author={Xu, Chuan and Wang, Libin and Liu, Zhibo and Chen, Long and Guo, Jingkun and Kang, Ning and Ma, Xiu-Liang and Cheng, Hui-Ming and Ren, Wencai},
  journal={Nature materials},
  volume={14},
  number={11},
  pages={1135--1141},
  year={2015},
  publisher={Nature Publishing Group UK London}
}

@article{ref:chan2024fabrication,
  title={The fabrication of Ti 3 C 2 and Ti 3 CN MXenes by electrochemical etching},
  author={Chan, Kai Chio and Guan, Xiang and Zhang, Teng and Lin, Kailing and Huang, Yihe and Lei, Lingshu and Georgantas, Yiannis and Gogotsi, Yury and Bissett, Mark A and Kinloch, Ian A},
  journal={Journal of Materials Chemistry A},
  volume={12},
  number={37},
  pages={25165--25175},
  year={2024},
  publisher={Royal Society of Chemistry}
}

@article{ref:amrillah2022towards,
  title={Towards greener and more sustainable synthesis of MXenes: a review},
  author={Amrillah, Tahta and Abdullah, Che Azurahanim Che and Hermawan, Angga and Sari, Fitri Nur Indah and Alviani, Vani Novita},
  journal={Nanomaterials},
  volume={12},
  number={23},
  pages={4280},
  year={2022},
  publisher={MDPI}
}

@article{ref:wyatt2024alkali,
  title={Alkali cation stabilization of defects in 2D MXenes at ambient and elevated temperatures},
  author={Wyatt, Brian C and Boebinger, Matthew G and Hood, Zachary D and Adhikari, Shiba and Micha{\l}owski, Pawe{\l} Piotr and Nemani, Srinivasa Kartik and Muraleedharan, Murali Gopal and Bedford, Annabelle and Highland, Wyatt J and Kent, Paul RC and others},
  journal={Nature Communications},
  volume={15},
  number={1},
  pages={6353},
  year={2024},
  publisher={Nature Publishing Group UK London}
}

@article{ref:tao2019atomically,
  title={Atomically layered and ordered rare-earth i-MAX phases: a new class of magnetic quaternary compounds},
  author={Tao, Quanzheng and Lu, Jun and Dahlqvist, Martin and Mockute, Aurelija and Calder, Stuart and Petruhins, Andrejs and Meshkian, Rahele and Rivin, Oleg and Potashnikov, Daniel and Caspi, El’ad N and others},
  journal={Chemistry of Materials},
  volume={31},
  number={7},
  pages={2476--2485},
  year={2019},
  publisher={ACS Publications}
}

@article{ref:novoselov20162d,
  title={2D materials and van der Waals heterostructures},
  author={Novoselov, K Sꎬ and Mishchenko, Artem and Carvalho, Alexandra and Castro Neto, AH},
  journal={Science},
  volume={353},
  number={6298},
  pages={aac9439},
  year={2016},
  publisher={American Association for the Advancement of Science}
}

@article{ref:autere2018nonlinear,
  title={Nonlinear optics with 2D layered materials},
  author={Autere, Anton and Jussila, Henri and Dai, Yunyun and Wang, Yadong and Lipsanen, Harri and Sun, Zhipei},
  journal={Advanced Materials},
  volume={30},
  number={24},
  pages={1705963},
  year={2018},
  publisher={Wiley Online Library}
}

@article{ref:sun2010graphene,
  title={Graphene mode-locked ultrafast laser},
  author={Sun, Zhipei and Hasan, Tawfique and Torrisi, Felice and Popa, Daniel and Privitera, Giulia and Wang, Fengqiu and Bonaccorso, Francesco and Basko, Denis M and Ferrari, Andrea C},
  journal={ACS nano},
  volume={4},
  number={2},
  pages={803--810},
  year={2010},
  publisher={ACS Publications}
}

@article{ref:vermeulen2022perspectives,
  title={Perspectives on nonlinear optics of graphene: Opportunities and challenges},
  author={Vermeulen, Nathalie},
  journal={APL Photonics},
  volume={7},
  number={2},
  year={2022},
  publisher={AIP Publishing}
}

@article{ref:gu2012regenerative,
  title={Regenerative oscillation and four-wave mixing in graphene optoelectronics},
  author={Gu, Tingyi and Petrone, Nick and McMillan, James F and van der Zande, Arend and Yu, Mingbin and Lo, Guo-Qiang and Kwong, Dim-Lee and Hone, James and Wong, Chee Wei},
  journal={Nature photonics},
  volume={6},
  number={8},
  pages={554--559},
  year={2012},
  publisher={Nature Publishing Group UK London}
}

@article{ref:wen2019nonlinear,
  title={Nonlinear optics of two-dimensional transition metal dichalcogenides},
  author={Wen, Xinglin and Gong, Zibo and Li, Dehui},
  journal={InfoMat},
  volume={1},
  number={3},
  pages={317--337},
  year={2019},
  publisher={Wiley Online Library}
}

@article{ref:zheng2015characterization,
  title={Characterization of nonlinear properties of black phosphorus nanoplatelets with femtosecond pulsed Z-scan measurements},
  author={Zheng, Xin and Chen, Runze and Shi, Gang and Zhang, Jianwei and Xu, Zhongjie and Cheng, Xiang’ai and Jiang, Tian},
  journal={Optics letters},
  volume={40},
  number={15},
  pages={3480--3483},
  year={2015},
  publisher={Optical Society of America}
}

@article{ref:abdelwahab2019giant,
  title={Giant and tunable optical nonlinearity in single-crystalline 2D perovskites due to excitonic and plasma effects},
  author={Abdelwahab, Ibrahim and Dichtl, Paul and Grinblat, Gustavo and Leng, Kai and Chi, Xiao and Park, In-Hyeok and Nielsen, Michael P and Oulton, Rupert F and Loh, Kian Ping and Maier, Stefan A},
  journal={Advanced Materials},
  volume={31},
  number={29},
  pages={1902685},
  year={2019},
  publisher={Wiley Online Library}
}

@article{ref:shang2022nonlinear,
  title={Nonlinear optical properties of MXene and applications in broadband ultrafast photonics},
  author={Shang, Ce and Zhang, Yifan and Wang, Gang and Sun, Jingxuan and Cheng, Yuan and Zhang, Yong-Biao and Yao, Baicheng and Fu, Bo and Li, Jiebo},
  journal={Journal of Alloys and Compounds},
  volume={918},
  pages={165580},
  year={2022},
  publisher={Elsevier}
}

@article{ref:hazan2023mxene,
  title={MXene-nanoflakes-enabled all-optical nonlinear activation function for on-chip photonic deep neural networks},
  author={Hazan, Adir and Ratzker, Barak and Zhang, Danzhen and Katiyi, Aviad and Sokol, Maxim and Gogotsi, Yury and Karabchevsky, Alina},
  journal={Advanced materials},
  volume={35},
  number={11},
  pages={2210216},
  year={2023},
  publisher={Wiley Online Library}
}

@article{ref:jiang2018broadband,
  title={Broadband nonlinear photonics in few-layer MXene Ti3C2Tx (T= F, O, or OH)},
  author={Jiang, Xiantao and Liu, Shunxiang and Liang, Weiyuan and Luo, Shaojuan and He, Zhiliang and Ge, Yanqi and Wang, Huide and Cao, Rui and Zhang, Feng and Wen, Qiao and others},
  journal={Laser \& photonics reviews},
  volume={12},
  number={2},
  pages={1700229},
  year={2018},
  publisher={Wiley Online Library}
}

@article{ref:park2024ultrahigh,
  title={Ultrahigh Nonlinear Responses from MXene Plasmons in the Short-Wave Infrared Range},
  author={Park, Changhoon and Park, Nu-Ri and Kwon, Jisung and Kim, Hyerim and Gogotsi, Yury and Koo, Chong Min and Kim, Myung-Ki},
  journal={Advanced Materials},
  volume={36},
  number={21},
  pages={2309189},
  year={2024},
  publisher={Wiley Online Library}
}

@article{ref:zhang2021third,
  title={Third-order nonlinear properties and reverse absorption behavior in layered Ti3C2 MXene in the near infrared region},
  author={Zhang, Tong and Chu, Hongwei and Li, Ying and Zhao, Shengzhi and Li, Dechun},
  journal={Optical Materials Express},
  volume={11},
  number={12},
  pages={4051--4057},
  year={2021},
  publisher={Optica Publishing Group}
}

@article{ref:li2020broadband,
  title={Broadband spatial self-phase modulation and ultrafast response of MXene Ti3C2Tx (T= O, OH or F)},
  author={Li, Jie and Zhang, Zilong and Yi, Jun and Miao, Lili and Huang, Jing and Zhang, Jinrui and He, Yuan and Huang, Bin and Zhao, Chujun and Zou, Yanhong and others},
  journal={Nanophotonics},
  volume={9},
  number={8},
  pages={2415--2424},
  year={2020},
  publisher={De Gruyter}
}

@article{ref:lee2021investigation,
  title={Investigation on the nonlinear optical properties of V 2 C MXene at 1.9 $\mu$m},
  author={Lee, Jinho and Kwon, Suh-Young and Lee, Ju Han},
  journal={Journal of Materials Chemistry C},
  volume={9},
  number={42},
  pages={15346--15353},
  year={2021},
  publisher={Royal Society of Chemistry}
}

@article{ref:lee2023nonlinear,
  title={Nonlinear absorption and refraction properties of V4C3 MXene and its use for an ultra-broadband saturable absorber},
  author={Lee, Kyungtaek and Kwon, Suh-young and Woo, Taeho and Ryu, Janghyun and Jung, Junha and Lee, Ju Han},
  journal={Advanced Optical Materials},
  volume={11},
  number={13},
  pages={2300213},
  year={2023},
  publisher={Wiley Online Library}
}

@article{reF:xiao2021near,
  title={Near-infrared spatial self-phase modulation in ultrathin niobium carbide nanosheets},
  author={Xiao, Si and He, Yi-lin and Dong, Yu-lan and Wang, Yi-duo and Zhou, Li and Zhang, Xue-jun and Wang, Ying-wei and He, Jun},
  journal={Frontiers in Physics},
  volume={9},
  pages={674820},
  year={2021},
  publisher={Frontiers Media SA}
}

@inproceedings{ref:lee2023investigation,
    author = {Kyungtaek Lee and Suh-Young Kwon and Jaehak Choi and Janghyun Ryu and Taeho Woo and Junha Jung and Ju Han Lee},
    booktitle = {Frontiers in Optics $+$ Laser Science 2023 (FiO, LS)},
    journal = {Frontiers in Optics $+$ Laser Science 2023 (FiO, LS)},
    pages = {JTu5A.39},
    publisher = {Optica Publishing Group},
    title = {Investigation of Nonlinear Absorption and Nonlinear Refraction Properties of Nb4C3 MXene at 1900 nm},
    year = {2023},
    url = {https://opg.optica.org/abstract.cfm?URI=LS-2023-JTu5A.39},
    doi = {10.1364/FIO.2023.JTu5A.39},
}

@article{ref:zhao2024two,
  title={Two-Dimensional Nb4AlC3 (MAX) and Nb4C3 (MXene): Excellent Optical Limiting Materials with Oppositely Applicated Waveband},
  author={Zhao, Zhenyu and Du, Binjian and Ren, Zihan and Liu, Qianhou and Xing, Fei and Zhang, Fang},
  journal={ACS Applied Nano Materials},
  volume={7},
  number={3},
  pages={2775--2785},
  year={2024},
  publisher={ACS Publications}
}

@article{ref:jin2024thickness,
  title={Thickness-and Wavelength-Dependent Nonlinear Optical Absorption in 2D Layered MXene Films},
  author={Jin, Di and Liu, Wenbo and Jia, Linnan and Zhang, Yuning and Hu, Junkai and El Dirani, Houssein and Kerdiles, S{\'e}bastien and Sciancalepore, Corrado and Demongodin, Pierre and Grillet, Christian and others},
  journal={Small Science},
  volume={4},
  number={8},
  pages={2400179},
  year={2024},
  publisher={Wiley Online Library}
}

@article{ref:wang20232d,
  title={2D Nb 2 CT x MXene/MoS 2 heterostructure construction for nonlinear optical absorption modulation},
  author={Wang, Yiduo and Wang, Yingwei and Dong, Yulan and Zhou, Li and Kang, Jianlong and Wang, Ning and Li, Yejun and Yuan, Xiaoming and Zhang, Zhengwei and Huang, Han and others},
  journal={Opto-Electronic Advances},
  volume={6},
  number={10},
  pages={220162--1},
  year={2023}
}

@book{ref:svelto1998principles,
  title={Principles of lasers},
  author={Svelto, Orazio and Hanna, David C and others},
  volume={4},
  year={1998},
  publisher={Springer}
}

@article{ref:pan2007passively,
  title={Passively $Q$-switched Ytterbium-Doped Double-Clad Fiber Laser With a Cr$^{4+}$:YAG Saturable Absorber},
  author={Pan, Lei and Utkin, Ilya and Fedosejevs, Robert},
  journal={IEEE Photonics technology letters},
  volume={19},
  number={24},
  pages={1979--1981},
  year={2007},
  publisher={IEEE}
}

@article{ref:huang2009high,
  title={High-pulse-energy, passively Q-switched Yb-doped fiber laser with AlGaInAs quantum wells as a saturable absorber},
  author={Huang, JY and Huang, WC and Zhuang, WZ and Su, Kuan-Wei and Chen, Yung-Fu and Huang, Kai-Feng},
  journal={Optics letters},
  volume={34},
  number={15},
  pages={2360--2362},
  year={2009},
  publisher={Optical Society of America}
}

@article{ref:zhou2009tunable,
  title={Tunable passively $Q$-switched erbium-doped fiber laser with carbon nanotubes as a saturable absorber},
  author={Zhou, Da-Peng and Wei, Li and Dong, Bo and Liu, Wing-Ki},
  journal={IEEE Photonics Technology Letters},
  volume={22},
  number={1},
  pages={9--11},
  year={2009},
  publisher={IEEE}
}

@article{ref:liu2013carbon,
  title={Carbon-nanotube-based passively Q-switched fiber laser for high energy pulse generation},
  author={Liu, HH and Chow, KK and Yamashita, Shinji and Set, SY},
  journal={Optics \& laser technology},
  volume={45},
  pages={713--716},
  year={2013},
  publisher={Elsevier}
}

@article{ref:li2023nanometer,
  title={Nanometer-Thin Stacks of MXenes (Ti3C2T x and Ta4C3T x) for Applications as Nonlinear Photonic Devices},
  author={Li, Genglin and Du, Wenhui and Sun, Shuo and Chen, Zhixiang and Liu, Hongliang and Lu, Qingming and Sun, Xiaoli and Ma, Yandong and Jia, Yuechen and Chen, Feng},
  journal={ACS Applied Nano Materials},
  volume={6},
  number={2},
  pages={856--866},
  year={2023},
  publisher={ACS Publications}
}

@article{ref:tian2020mxene,
  title={MXene Ti3C2Tx saturable absorber for passively Q-switched mid-infrared laser operation of femtosecond-laser--inscribed Er: Y2O3 ceramic channel waveguide},
  author={Tian, Qingyu and Yin, Peng and Zhang, Teng and Zhou, Lunbin and Xu, Bin and Luo, Zhengqian and Liu, Hongliang and Ge, Yanqi and Zhang, Jian and Liu, Peng and others},
  journal={Nanophotonics},
  volume={9},
  number={8},
  pages={2495--2503},
  year={2020},
  publisher={De Gruyter}
}

@article{ref:thomas2023review,
  title={A review of Nb2CT x MXene as an emerging 2D material: synthesis, applications in rechargeable batteries and supercapacitors, progress, and outlook},
  author={Thomas, Susmi Anna and Cherusseri, Jayesh},
  journal={Energy \& Fuels},
  volume={37},
  number={11},
  pages={7555--7576},
  year={2023},
  publisher={ACS Publications}
}

@article{ref:mathis2021modified,
  title={Modified MAX phase synthesis for environmentally stable and highly conductive Ti3C2 MXene},
  author={Mathis, Tyler S and Maleski, Kathleen and Goad, Adam and Sarycheva, Asia and Anayee, Mark and Foucher, Alexandre C and Hantanasirisakul, Kanit and Shuck, Christopher E and Stach, Eric A and Gogotsi, Yury},
  journal={ACS nano},
  volume={15},
  number={4},
  pages={6420--6429},
  year={2021},
  publisher={ACS Publications}
}

@article{ref:matthews2021guidelines,
  title={Guidelines for synthesis and processing of chemically stable two-dimensional V2CT x MXene},
  author={Matthews, Kyle and Zhang, Teng and Shuck, Christopher E and VahidMohammadi, Armin and Gogotsi, Yury},
  journal={Chemistry of Materials},
  volume={34},
  number={2},
  pages={499--509},
  year={2021},
  publisher={ACS Publications}
}

@article{ref:zhu2024additive,
  title={Additive-free anode with high stability: Nb2CT x MXene prepared by HCl-LiF hydrothermal etching for Lithium-ion batteries},
  author={Zhu, Xiaoxue and Yang, Kai and Zhang, Zhen and He, Siyuan and Shen, Zihao and Jiang, Wei and Huang, Yiling and Xu, Yan and Jiang, Qiutong and Pan, Limei and others},
  journal={ACS Applied Materials \& Interfaces},
  volume={16},
  number={22},
  pages={28709--28718},
  year={2024},
  publisher={ACS Publications}
}

@article{ref:minassian2025nonlinear,
  title={Nonlinear light absorption in Ti 3 C 2 T x MXene: a theoretical study},
  author={Minassian, Hayk and Melikyan, Armen and Gon{\c{c}}alves, Manuel Rodrigues and Petrosyan, Petros},
  journal={RSC advances},
  volume={15},
  number={29},
  pages={23934--23942},
  year={2025},
  publisher={Royal Society of Chemistry}
}

@article{ref:tang2012mxenes,
  title={Are MXenes promising anode materials for Li ion batteries? Computational studies on electronic properties and Li storage capability of Ti3C2 and Ti3C2X2 (X= F, OH) monolayer},
  author={Tang, Qing and Zhou, Zhen and Shen, Panwen},
  journal={Journal of the American Chemical Society},
  volume={134},
  number={40},
  pages={16909--16916},
  year={2012},
  publisher={ACS Publications}
}

@article{ref:saha2023surface,
  title={Surface-modified MXenes: Simulation to potential applications},
  author={Saha, Dipankar and Dalmieda, Johnson and Patel, Vinay},
  journal={ACS Applied Electronic Materials},
  volume={5},
  number={6},
  pages={2933--2955},
  year={2023},
  publisher={ACS Publications}
}

@inproceedings{ref:kim2018femtosecond,
  title={Femtosecond mode-locked fiber laser using an etched optical fiber immersed in liquid Mxene as a saturable absorber},
  author={Kim, Jeong Je and Kim, Hyerim and Koo, Chong Min and Lee, Jae Ha and Lee, Sang Bae and Lee, Kwanil},
  booktitle={2018 23rd Opto-Electronics and Communications Conference (OECC)},
  pages={1--2},
  year={2018},
  organization={IEEE}
}

@article{ref:jhon2017metallic,
  title={Metallic MXene saturable absorber for femtosecond mode-locked lasers},
  author={Jhon, Young In and Koo, Joonhoi and Anasori, Babak and Seo, Minah and Lee, Ju Han and Gogotsi, Yury and Jhon, Young Min},
  journal={Advanced Materials},
  volume={29},
  number={40},
  pages={1702496},
  year={2017},
  publisher={Wiley Online Library}
}

@article{ref:dong2018saturable,
  title={Saturable absorption in 2D Ti3C2 MXene thin films for passive photonic diodes},
  author={Dong, Yongchang and Chertopalov, Sergii and Maleski, Kathleen and Anasori, Babak and Hu, Longyu and Bhattacharya, Sriparna and Rao, Apparao M and Gogotsi, Yury and Mochalin, Vadym N and Podila, Ramakrishna},
  journal={Advanced materials},
  volume={30},
  number={10},
  pages={1705714},
  year={2018},
  publisher={Wiley Online Library}
}

@article{ref:wang2020mxene,
  title={MXene photonic devices for near-infrared to mid-infrared ultrashort pulse generation},
  author={Wang, Zhenhong and Li, Hongbo and Luo, Mulin and Chen, Tenghui and Xia, Xuefeng and Chen, Hualong and Ma, Chunyang and Guo, Jia and He, Zhenwu and Song, Yufeng and others},
  journal={ACS applied nano materials},
  volume={3},
  number={4},
  pages={3513--3522},
  year={2020},
  publisher={ACS Publications}
}

@article{ref:rosol2024short,
  title={Short pulsed (1088.5 nm) neodymium-doped fiber laser via Ti3C2TxPVA film-based SA},
  author={Rosol, AHA and Hamzah, A and Zulkipli, NF and Harun, SW},
  journal={Optik},
  volume={296},
  pages={171552},
  year={2024},
  publisher={Elsevier}
}

@article{ref:wang2024high,
  title={A high stability and low noise passively Q-switched yellow-green laser with a Ti3C2Tx-PVA saturable absorber},
  author={Wang, Guozhen and Bai, Yang and Li, Yi and Yan, Jiale and Yang, Shunyu and Lu, Baole and Zhang, Heng},
  journal={Optics Communications},
  volume={560},
  pages={130473},
  year={2024},
  publisher={Elsevier}
}

@article{ref:liang2023passively,
  title={Passively Q-switched erbium-doped fiber laser based on Ti3C2Tx saturable absorber},
  author={Liang, Lei and Cheng, Jiawei and Liu, Nan and Zhang, Jinniu and Ren, Kaili and Zhao, Qiyi and Li, Lu},
  journal={Optik},
  volume={286},
  pages={171045},
  year={2023},
  publisher={Elsevier}
}

@article{ref:qian2025recent,
  title={Recent advances in 2D MXene-based heterostructures for gas sensing: mechanisms and applications in environmental and biomedical fields},
  author={Qian, Lanting and Rahmati, Farnood and Li, Fengchao and Zhang, Tianzhu and Wang, Tao and Zhang, Haoze and Yan, Shuo and Zheng, Yun},
  journal={Nanoscale},
  year={2025},
  publisher={Royal Society of Chemistry}
}

@article{ref:lu20212d,
  title={2D MXene nanomaterials for versatile biomedical applications: current trends and future prospects},
  author={Lu, Beibei and Zhu, Zhenye and Ma, Biyuan and Wang, Wei and Zhu, Rongshu and Zhang, Jiaheng},
  journal={Small},
  volume={17},
  number={46},
  pages={2100946},
  year={2021},
  publisher={Wiley Online Library}
}

@article{ref:kim2018metallic,
  title={Metallic Ti3C2T x MXene gas sensors with ultrahigh signal-to-noise ratio},
  author={Kim, Seon Joon and Koh, Hyeong-Jun and Ren, Chang E and Kwon, Ohmin and Maleski, Kathleen and Cho, Soo-Yeon and Anasori, Babak and Kim, Choong-Ki and Choi, Yang-Kyu and Kim, Jihan and others},
  journal={ACS nano},
  volume={12},
  number={2},
  pages={986--993},
  year={2018},
  publisher={ACS Publications}
}

@article{ref:john2023gas,
  title={Gas-sensing mechanisms and performances of MXenes and MXene-based heterostructures},
  author={John, Riya Alice B and Vijayan, Karthikeyan and Septiani, Ni Luh Wulan and Hardiansyah, Andri and Kumar, A Ruban and Yuliarto, Brian and Hermawan, Angga},
  journal={Sensors},
  volume={23},
  number={21},
  pages={8674},
  year={2023},
  publisher={MDPI}
}

@article{ref:shin2024high,
  title={High-performance H 2 S gas sensor utilizing MXene/MoS 2 heterostructure synthesized via the Langmuir--Blodgett technique and chemical vapor deposition},
  author={Shin, Jae Hyuk and Jo, Su Hun and Rhyu, Hyejin and Park, Chanwon and Kang, Myung Hyun and Song, Wooseok and Lee, Sun Sook and Lim, Jongsun and Myung, Sung},
  journal={RSC advances},
  volume={14},
  number={51},
  pages={37781--37787},
  year={2024},
  publisher={Royal Society of Chemistry}
}

@article{ref:wu2018few,
  title={Few-layer Ti3C2Tx MXene: A promising surface plasmon resonance biosensing material to enhance the sensitivity},
  author={Wu, Leiming and You, Qi and Shan, Youxian and Gan, Shuaiwen and Zhao, Yuting and Dai, Xiaoyu and Xiang, Yuanjiang},
  journal={Sensors and Actuators B: Chemical},
  volume={277},
  pages={210--215},
  year={2018},
  publisher={Elsevier}
}

@article{ref:sharma2025bound,
  title={Bound-state-in-the-continuum (BIC) induced narrow resonances in MXene-coated absorptive dielectric metasurfaces for Methane sensing},
  author={Sharma, Shubhanshi and Pradhan, Monica and Katiyi, Aviad and Karabchevsky, Alina and Varshney, Shailendra Kumar},
  journal={arXiv preprint arXiv:2509.04616},
  year={2025}
}

@article{ref:guo2024mxene,
  title={MXene/WO3 Sensor Array with Improved SNN Algorithm for Accurate Identification of Toxic Gases},
  author={Guo, Liangchao and Wang, Junke and Han, Haoran and Wang, Peng and Lu, Yunxiang and Yuan, Qilong and Du, Chunyu and Yin, Shuo and Zhou, Ye and Zhang, Chao},
  journal={ACS Applied Materials \& Interfaces},
  volume={16},
  number={45},
  pages={62421--62428},
  year={2024},
  publisher={ACS Publications}
}

@article{ref:duport2012all,
  title={All-optical reservoir computing},
  author={Duport, Fran{\c{c}}ois and Schneider, Bendix and Smerieri, Anteo and Haelterman, Marc and Massar, Serge},
  journal={Optics express},
  volume={20},
  number={20},
  pages={22783--22795},
  year={2012},
  publisher={Optical Society of America}
}

@article{ref:zou2008overview,
  title={Overview of artificial neural networks},
  author={Zou, Jinming and Han, Yi and So, Sung-Sau},
  journal={Artificial neural networks: methods and applications},
  pages={14--22},
  year={2008},
  publisher={Springer}
}

@article{ref:shastri2021photonics,
  title={Photonics for artificial intelligence and neuromorphic computing},
  author={Shastri, Bhavin J and Tait, Alexander N and Ferreira de Lima, Thomas and Pernice, Wolfram HP and Bhaskaran, Harish and Wright, C David and Prucnal, Paul R},
  journal={Nature Photonics},
  volume={15},
  number={2},
  pages={102--114},
  year={2021},
  publisher={Nature Publishing Group UK London}
}

@article{ref:yang2022mxene,
  title={MXene-based broadband ultrafast nonlinear activator for optical computing},
  author={Yang, Zhan and Tan, Weimin and Zhang, Tianju and Chen, Chenduan and Wang, Zixin and Mao, Yu and Ma, Chenxi and Lin, Qing and Bi, Wanjun and Yu, Fei and others},
  journal={Advanced Optical Materials},
  volume={10},
  number={17},
  pages={2200714},
  year={2022},
  publisher={Wiley Online Library}
}

@article{ref:stavrou2025emerging,
  title={Emerging Ta4C3 and Mo2Ti2C3 MXene Nanosheets for Ultrafast Photonics},
  author={Stavrou, Michalis and Chacon, Benjamin and Farsari, Maria and Pappa, Anna Maria and Delogu, Lucia Gemma and Gogotsi, Yury and Gray, David},
  journal={Advanced Optical Materials},
  pages={2403277},
  year={2025},
  publisher={Wiley Online Library}
}

@article{ref:debow2024two,
  title={Two-Dimensional MXene Flakes with Large Second Harmonic Generation and Unique Surface Responses},
  author={Debow, Shaun M and Fisher, Haley and Brown, Jesse B and Liebes, Mallory and Wang, Hui and Phan, TT Trinh and Mac, T Kien and Trinh, M Tuan and Su, Yanqing and Zander, Zachary and others},
  journal={The Journal of Physical Chemistry Letters},
  volume={15},
  number={44},
  pages={11087--11096},
  year={2024},
  publisher={ACS Publications}
}

@article{ref:he2020strain,
  title={Strain-dependent anisotropic nonlinear optical response in two-dimensional functionalized MXene Sc 2 CT 2 (T= O and OH)},
  author={He, Chuan and Zhao, Qiyi and Huang, Yuanyuan and Du, Wanyi and Zhu, Lipeng and Zhou, Yixuan and Zhang, Sujuan and Xu, Xinlong},
  journal={Physical Chemistry Chemical Physics},
  volume={22},
  number={37},
  pages={21428--21435},
  year={2020},
  publisher={Royal Society of Chemistry}
}

@article{ref:huang2020highly,
  title={Highly stable MXene (V2CTx)-based harmonic pulse generation},
  author={Huang, Weichun and Ma, Chunyang and Li, Chao and Zhang, Ye and Hu, Lanping and Chen, Tingting and Tang, Yanfeng and Ju, Jianfeng and Zhang, Han},
  journal={Nanophotonics},
  volume={9},
  number={8},
  pages={2577--2585},
  year={2020},
  publisher={De Gruyter}
}

@article{ref:uddin2024asymmetric,
  title={Asymmetric Laser Field Interaction with MXene Coated on the Side Surface of Optical Fibers for Ultrafast Nonlinear Switches},
  author={Uddin, Siam and Debnath, Pulak C and Kim, Hyerim and Moon, Hyowon and Koo, Chong Min and Song, Yong-Won},
  journal={ACS Applied Materials \& Interfaces},
  volume={16},
  number={7},
  pages={9137--9143},
  year={2024},
  publisher={ACS Publications}
}

@article{ref:wang2020two,
  title={Two-photon absorption in monolayer MXenes},
  author={Wang, Gaozhong and Bennett, Daniel and Zhang, Chuanfang and {\'O} Coile{\'a}in, Cormac and Liang, Meiying and McEvoy, Niall and Wang, Jing Jing and Wang, Jun and Wang, Kangpeng and Nicolosi, Valeria and others},
  journal={Advanced Optical Materials},
  volume={8},
  number={9},
  pages={1902021},
  year={2020},
  publisher={Wiley Online Library}
}

@article{ref:ge2026all,
  title={All-optical logic processing unit using Kerr nonlinearity of MXene},
  author={Ge, Yanqi and Wang, Wenkai and Wang, Meng and Zhang, Shoukai and Zhong, Yuanjiang and Zhang, Yule and Wei, Songrui and Du, Bowen and Wang, Huide and Qiu, Meng and others},
  journal={Nature Communications},
  year={2026},
  publisher={Nature Publishing Group UK London}
}

@article{ref:fang2026band,
  title={Band-like Optical Signatures of Ti3C2T x MXenes},
  author={Fang, Hui and Fang, Zhenyao and Thakur, Anupma and Rad, Vahid and Chandran Balachandran Sajitha, Nithin and Micha{\l}owski, Pawe{\l} P and Soroush, Masoud and Anasori, Babak and Rappe, Andrew M and Fakhraai, Zahra},
  journal={The Journal of Physical Chemistry C},
  year={2026},
  publisher={ACS Publications}
}

@article{ref:stavrou2026slow,
  title={Slow light-matter interactions in MXene-integrated photonic crystals in the telecommunications spectrum},
  author={Stavrou, Michalis and Matthaiakakis, Nikolaos and Kambouraki, Elmina and Chacon, Benjamin and {\v{C}}eponkus, Justinas and Malinauskas, Mangirdas and Juodkazis, Saulius and Loukakos, Panagiotis and Gray, David and Gogotsi, Yury and others},
  year={2026}
}

@article{ref:qin2025enhanced,
  title={Enhanced third-harmonic generation and degenerate four-wave mixing in an all-dielectric metasurface via Brillouin zone folding induced bound states in the continuum},
  author={Qin, Meibao and Wu, Feng and Liu, Tingting and Zhang, Dandan and Xiao, Shuyuan},
  journal={Physical Review B},
  volume={111},
  number={3},
  pages={035414},
  year={2025},
  publisher={APS}
}

@article{ref:zeng2023mxene,
  title={MXene-based memristor for artificial optoelectronic neuron},
  author={Zeng, Bangyan and Zhang, Xianghong and Gao, Changsong and Zou, Yi and Yu, Xipeng and Yang, Qian and Guo, Tailiang and Chen, Huipeng},
  journal={IEEE Transactions on Electron Devices},
  volume={70},
  number={3},
  pages={1359--1365},
  year={2023},
  publisher={IEEE}
}

@article{ref:feng2023novel,
  title={A novel nonvolatile memory device based on oxidized Ti3C2Tx MXene for neurocomputing application},
  author={Feng, Xin and Huang, Jingjing and Ning, Jing and Wang, Dong and Zhang, Jincheng and Hao, Yue},
  journal={Carbon},
  volume={205},
  pages={365--372},
  year={2023},
  publisher={Elsevier}
}

@article{ref:huang2023flexible,
  title={Flexible, transparent, and wafer-scale artificial synapse array based on TiOx/Ti3C2Tx film for neuromorphic computing},
  author={Huang, Junhua and Yang, Shaodian and Tang, Xin and Yang, Leilei and Chen, Wenjun and Chen, Zibo and Li, Xinming and Zeng, Zhiping and Tang, Zikang and Gui, Xuchun},
  journal={Advanced Materials},
  volume={35},
  number={33},
  pages={2303737},
  year={2023},
  publisher={Wiley Online Library}
}

@article{ref:wang2024dual,
  title={Dual biological-clock controllable low-power fibrous synapse array based on heterojunction switched conductive filaments},
  author={Wang, Kaiyang and Ren, Shuhui and Jia, Yunfang and Yan, Xiaobing},
  journal={Nano Energy},
  volume={127},
  pages={109765},
  year={2024},
  publisher={Elsevier}
}

@article{ref:aadhi2025scalable,
  title={Scalable photonic reservoir computing for parallel machine learning tasks},
  author={Aadhi, A and Di Lauro, L and Fischer, B and Dmitriev, P and Alamgir, I and Mazoukh, C and Perron, N and Viktorov, EA and Kovalev, AV and Eshaghi, A and others},
  journal={Nature Communications},
  year={2025},
  publisher={Nature Publishing Group UK London}
}

@article{ref:adeshina2025exploring,
  title={Exploring the frontier: nonlinear optics in low dimensional materials},
  author={Adeshina, Mohammad A and Kim, Hyunmin},
  journal={Nanophotonics},
  volume={14},
  number={10},
  pages={1451--1473},
  year={2025},
  publisher={De Gruyter}
}

@article{ref:simon2025tailoring,
  title={Tailoring optical response of MXene thin films},
  author={Simon, Jeffrey and Choi, Kyu Ri and Ippolito, Stefano and Prokopeva, Ludmila and Fruhling, Colton and Shalaev, Vladimir M and Kildishev, Alexander V and Gogotsi, Yury and Boltasseva, Alexandra},
  journal={Nanophotonics},
  volume={14},
  number={23},
  pages={4091--4097},
  year={2025},
  publisher={De Gruyter}
}

\end{document}